\documentclass{emulateapj}
\pdfoutput=1

\nonstopmode 
\usepackage[backref,breaklinks,colorlinks,citecolor=blue]{hyperref}
\usepackage[all]{hypcap} 

\usepackage[usenames,dvipsnames,svgnames,table]{xcolor}
\usepackage{enumerate}
\newcommand{\kms}{$\rm{kms^{-1}}$}
\newcommand{\pcm}{$\rm{cm^{-2}}$}
\providecommand{\CIV}{\ensuremath{\mbox{\ion{C}{4}}}}

\providecommand{\SiIV}{\ensuremath{\mbox{\ion{Si}{4}}}}

\providecommand{\NV}{\ensuremath{\mbox{\ion{N}{5}}}}
\providecommand{\HI}{\ensuremath{\mbox{\ion{H}{1}}}}
\providecommand{\OVI}{\ensuremath{\mbox{\ion{O}{6}}}}
\providecommand{\OVII}{\ensuremath{\mbox{\ion{O}{7}}}}
\providecommand{\NeVIII}{\ensuremath{\mbox{\ion{Ne}{8}}}}

\providecommand{\MgX}{\ensuremath{\mbox{\ion{Mg}{10}}}}
\providecommand{\SiXII}{\ensuremath{\mbox{\ion{Mg}{12}}}}

\usepackage{xparse}
\usepackage{soul}
\usepackage[usenames,dvipsnames]{xcolor}

\makeatletter
\NewDocumentCommand{\sayw}{O{Green}O{Black}+m}
    {%
        \begingroup
        \setulcolor{#1}%
        \setul{-0.6ex}{1.4pt}%
        \def\SOUL@uleverysyllable{%
            \rlap{%
                \color{#2}\the\SOUL@syllable
                \SOUL@setkern\SOUL@charkern}%
            \SOUL@ulunderline{%
                \phantom{\the\SOUL@syllable}}%
        }%
        \ul{#3}%
        \endgroup
    }
\makeatother

\usepackage[T1]{fontenc}
\usepackage{ae,aecompl,amsmath}

\shorttitle{The formation and physical origin of highly ionized cooling gas }
\shortauthors{Bordoloi et al.}

\begin{document}
\title{The formation and physical origin of highly ionized cooling gas}
\author{Rongmon Bordoloi\altaffilmark{1,2}, 
Alexander Y. Wagner\altaffilmark{3}, 
Timothy M. Heckman\altaffilmark{4} \& 
Colin A. Norman\altaffilmark{4,5} 
}
\altaffiltext{1}{MIT-Kavli Center for Astrophysics and Space Research, 77 Massachusetts Avenue, Cambridge, MA, 02139, USA;\href{mailto:bordoloi@mit.edu}{bordoloi@mit.edu}}
\altaffiltext{2}{Hubble Fellow}
\altaffiltext{3}{University of Tsukuba, Center for Computational Sciences, Tennodai 1-1-1, Tsukuba, Ibaraki, Japan}
\altaffiltext{4}{Department of Physics and Astronomy, John Hopkins University, 21218, Baltimore, MD}
\altaffiltext{5}{Space Telescope Science Institute, 3700 San Martin Drive, 21218, Baltimore, MD}


\begin{abstract}
We present a simple model that explains the origin of warm diffuse gas seen primarily as highly ionized absorption line systems in the spectra of background sources. We predict the observed column densities of several highly ionized transitions such as {\OVI}, {\OVII}, {\NeVIII}, {\NV},  and {\MgX}; and present a unified comparison of the model predictions with absorption lines seen in the Milky Way disk, Milky Way halo, starburst galaxies, the circumgalactic medium and the intergalactic medium at low and high redshifts.  We show that diffuse gas seen in such diverse environments can be simultaneously explained by a simple model of radiatively cooling gas.  We show that most of such absorption line systems are consistent with being collisionally ionized, and estimate the maximum likelihood temperature of the gas in each observation. This model satisfactorily explains why {\OVI} is regularly observed around star-forming low-$z$ L* galaxies, and why {\NV} is rarely seen around the same galaxies. We further present some consequences of this model in quantifying the dynamics of the cooling gas around galaxies and predict the shock velocities associated with such flows.  A unique strength of this model is that while it has only one free (but physically well-constrained)  parameter, it nevertheless successfully reproduces the available data on O VI absorbers in the interstellar, circumgalactic, intra-group, and intergalactic media, as well as the available data on other absorption-line from highly ionized species.

\end{abstract}

\section{Introduction}
Quasar absorption line study of galaxies, groups, clusters, and the intergalactic medium (IGM) is one of the most versatile technique to detect gas which is otherwise too diffuse to be seen in emission. Our current picture of galaxy evolution consists of a galaxy sitting at the center of a dark matter halo and a diffuse and large reservoir of cool and warm gas that envelopes the galaxy. These halos of diffuse gas, aptly named the circumgalactic medium (CGM) harbor large reservoir of gas that drive galaxy evolution \citep{Tumlinson2011a,Tripp2011,Bordoloi2014c,Peeples2013a, Zhu_CaII2013}.

Large scale galaxy surveys \citep{York2000, Lilly2009_Article}, combined with the success of Ultraviolet spectroscopy with the Hubble Space Telescope (HST), have made it possible to systematically map the CGM and to correlate its properties with the host galaxy properties and the large-scale environment \citep{Tumlinson2013,Stocke2013,Bordoloi2011a,Chen2010a,Prochaska2011a, zhu2013a,Werk2013,Stocke14,Bordoloi2012a,Bordoloi2014c,Zhu2014,Borthakur2015,Borthakur2016}.  

Highly ionized gas in the CGM is particularly interesting. \cite{Tumlinson2011a} showed that the  presence of diffuse {\OVI} gas around a galaxy strongly depends on whether the host galaxy is star-forming or not. There are also recent observations of {\OVI} absorption detected in galaxy groups  \citep{Stocke14,Johnson2015a}. These observations enable us to put lower limits on the CGM metal mass budgets by assuming a conservative maximum ionization fractions for the highly ionized species used in such calculations \citep{Tumlinson2011a,Bordoloi2014c}. Knowing the true physical mechanism of the ionization of such gas would indeed reduce the uncertainties in these metal mass estimates. It was found that the CGM of L* galaxies contain at least as much Oxygen as in their ISM \citep{Tumlinson2011a}.  Similarly in the CGM of sub-L* galaxies, masses for Carbon in the CGM also likely exceed the masses in the ISM \citep{Bordoloi2014c}. Rare observations of diffuse {\NeVIII} gas around post-starburst galaxies suggest that ``warm-hot'' plasma might contain 10 to 150 times more mass than the cooler gas phases in post-starburst winds \citep{Tripp2011}.

Large absorption line surveys have utilized the Far Ultraviolet Spectroscopic Explorer (FUSE) telescope to study the hot diffuse ``coronal''  {\OVI} gas present  in the halo of the Milky Way \citep{Sembach2003, Fox2006}.  They found that such {\OVI} is almost ubiquitously detected in the halo of the Milky Way. \cite{Fox2006}  reported that 47 out of a total of 63 ``highly ionized'' absorbers are not associated with any 21 cm {\HI} High Velocity Cloud emission in the halo of the Milky Way. They further report that 29 out of 38 {\OVI} absorbers exhibit associated {\HI} absorption seen in the Lyman series.  Such observations have led many people to argue that the highly ionized gas may trace different phases of the diffuse halo gas, and ions such as {\OVI}, {\NeVIII}, {\NV} etc, might arise in collisionally-ionized gas \citep{Fox2006, Lehner2011, Fox2011}. \cite{Wakker2012} compared the observations of highly ionized gas {\OVI}, {\NV}, {\SiIV}, {\CIV} to different theoretical models. They compared the column density ratios of such species to different models including collisional ionization equilibrium \citep{Gnat2007}, shock ionization \citep{Dopita1996}, turbulent mixing layers \citep{Slavin1993}, conductive interfaces \citep{Borkowski1990}, static non-equilibrium ionization radiative cooling \citep{Edgar1986}, thick disk supernovae \citep{Shelton1998}, radiative cooling gas flows \citep{Benjamin1994} etc. \cite{Wakker2012} found that non-equilibrium ionization radiative cooling is important in producing such gas, but these models can not to explain the full suite of  observations. It is difficult to explain the ratio of {\NV} to {\OVI} column densities with a single density pure photoionization model \citep{Werk2016}.  Recent work by \cite{Thompson2016} has explored mechanics of the origin of the warm CGM gas as being in hot galactic winds cooling radiatively on large scales. But detailed comparison to observations have not been performed on such models. 

Observations of circumgalactic and intergalactic {\OVI} absorption at higher redshifts have made inferring the origin of such highly ionized absorption even more difficult. In CGM surveys such as COS-Halos \citep{Tumlinson2013,Werk2013}, it is seen that the  warm {\OVI} absorbers almost always have associated {\HI} absorption. The kinematics of such {\HI} gas suggest that they are typicality cooler $ <10^{5} K$ gas \citep{Tumlinson2013}. However, other observations have found that there are also rare examples where no {\HI} gas is observed that is associated with the observed {\OVI} absorption \citep{Stocke14}. Further, increasing numbers of warm {\OVI} absorbers  with associated broad {\HI} absorption are also being reported \citep{Tripp2001,Savage2011}. \cite{Savage2014} recently reported that 31\% (14/45) of their sample of {\OVI} absorbers lie in the temperature range of  5 < log T  < 6.

In this work, we employ a simple model introduced by \cite{Cooling_flow2002} that can explain the available observations simultaneously and which can give insight into the origin of the gas around galaxies. They argued that most of the observed {\OVI} absorption-line systems can be explained as collisionally ionized gas radiatively cooling behind a shock or other type of radiatively-cooling flow.  We stress that this cooling flow model is a generic description of how highly ionized absorbers originate and it doesn't make any assumptions about which astrophysical processes create the cooling flows or cooling shocks. There are plenty of astrophysical processes which are capable and indeed expected to generate these cooling flows. Such cooling flows may arise in galactic outflows or accretion flows into galactic halos as gas cools radiatively behind shocks or as gas collapses in cooling instabilities. The gas in the cooling flow cools radiatively and different highly ionized species such as {\OVI}, {\NeVIII} etc.   would be created along the cooling flow, in regions which are at temperatures at which the CIE ionic abundance of that species would allow it to exist. The line widths of these absorption lines depend on the cooling velocities in the region where that ionic species is created. Complex multiple absorption line systems along one line of sight would require several cooling layers or multiple shocks along that line of sight to explain its origin.

Our aims in this paper are (1) to explore the cooling flow model by \cite{Cooling_flow2002} in more detail by examining the flow conservation equations, and (2) to apply the cooling flow model to describe new and mostly Hubble Space Telescope ultraviolet observations of highly ionized {\OVI}, {\NeVIII}, {\NV} gas that has been performed in recent years, and constrain the origin of these highly ionized absorption line systems.

This paper is organized as follows. In Section 2 we describe the observations that are used to compare with the cooling flow model. In Section 3 we describe the theory of a cooling flow model for any radiatively cooling species. In section 4 we discuss how to compare this model to observations. In section 5 we compare this model with observations. In Section 6 we describe the dynamics of the cooling gas, and finally in section 7 we summarize our findings. 

\section{The Observations}
In this section we list all the observations that have been compiled from the literature to be used in this work. This compilation is not exhaustive and it is not meant to be a complete summary of all the observations to date. It is simply an extensive compilation of several works that represent observations covering a diverse range of environments: the disk of the Milky Way, the Milky Way halo, starburst galaxies, the low redshift IGM, the CGM around $z\approx 0.2$ L* galaxies, and the absorption lines detected around $z\approx$ 2--3 galaxies.

The primary measurements pertinent to our analysis are the column densities of various ionic species and their line widths (second moment of the line profiles). We use the Voigt profile column densities from the literature and the fitted Doppler b ($b_{D}$) parameters, which is translated to line widths as $\Delta v\;=\; 3b_{D}/\sqrt{2}$. We consider data that have been published from the following works: the {\OVI} data used for this work are collected from low and high redshift CGM studies of COS-Halos survey \citep{Tumlinson2011a}, the KODIAQ survey \citep{Lehner2014}, observations of starburst galaxies \cite{Grimes2009}, observations along the disk of the Milky Way \citep{Bowen2008}, and several IGM low and moderate redshift studies \citep{Tripp2008,Meiring2013,Stocke14,Savage2005,Burchett2015a,Pachat2017,Qu2016}.

The {\NeVIII} doublet is harder to observe, as it has extremely blue rest frame wavelengths of 770.4 {\AA} and 780.3 {\AA}. There are only handful of detection of {\NeVIII}, primarily in high S/N HST/COS or FUSE spectra presented in several publications \citep{Savage2005,Savage2011,Narayanan2009,Narayanan2011,Meiring2013,Hussain2015,Qu2016,Pachat2017}. We also include a new {\NeVIII} absorption detected towards QSO J1154+4635. This observation was done under HST PID (13852; PI Bordoloi). We detect {\OVI} along with {\NeVIII} and {\HI} along this sight line and the spectrum is show as Figure \ref{fig:appendix} in the Appendix. The measured {\OVI} column density and Doppler line width for this absorber are $\log\; N_{OVI}$/{\pcm} = 14.71$\pm$0.03, and $b_{OVI}$ = 99 $\pm$ 11 {\kms}. The measured {\NeVIII} column density and Doppler line width for this absorber are $\log\; N_{NeVIII}$/{\pcm} = 14.65$\pm$0.08, and $b_{NeVIII}$ = 187 $\pm$ 39 {\kms}. 

{\NV} is not commonly seen in the absorption line studies. We have compiled the handful of {\NV} observations from COS-Halos \citep{Werk2013}, high redshift {\NV} measurements associated with DLAs \citep{Fox2009}, low redshift IGM observations of of \citep{Savage2005,Burchett2015a} and the Milky Way HVC observations of \citep{Lehner2011}.

Lastly, we present measurements of {\OVII} absorption lines with high resolution \textit{Chandra} and \textit{XMM-Newton} spectra, that trace the warm and ionized gas of the Milky Way halo \citep{Gupta2012}. The {\OVII} column densities were estimated using the curve of growth analysis. The low velocity resolution of the spectra made it impossible to directly measure the {\OVII} line widths. Multiple absorption lines from the same ions were used to place limits on the column density and the Doppler $b$ parameters of the {\OVII} transitions \citep{Gupta2012}.

\begin{figure}
\includegraphics[height=7cm,width=9cm]{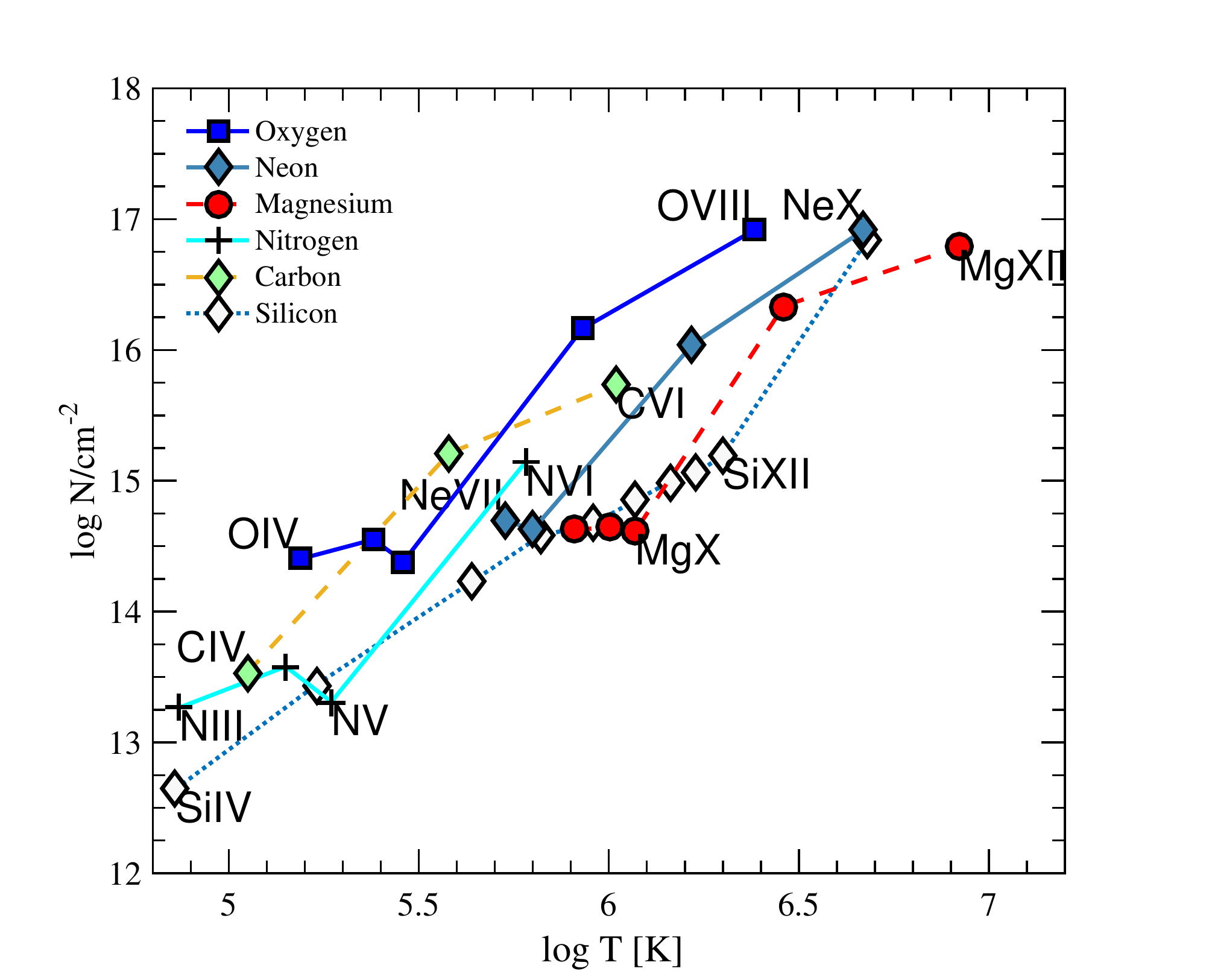}
  \caption{Model predictions for cooling flow column densities for several ions of abundant elements plotted at the temperature where each ion has its maximum fractional abundance assuming CIE. For simplicity, the model column densities are computed at the isothermal sound speed at that temperature.}
\label{fig:summary}
\end{figure}

\begin{figure*}
\centering
\includegraphics[height=6.5cm,width=7.8cm]{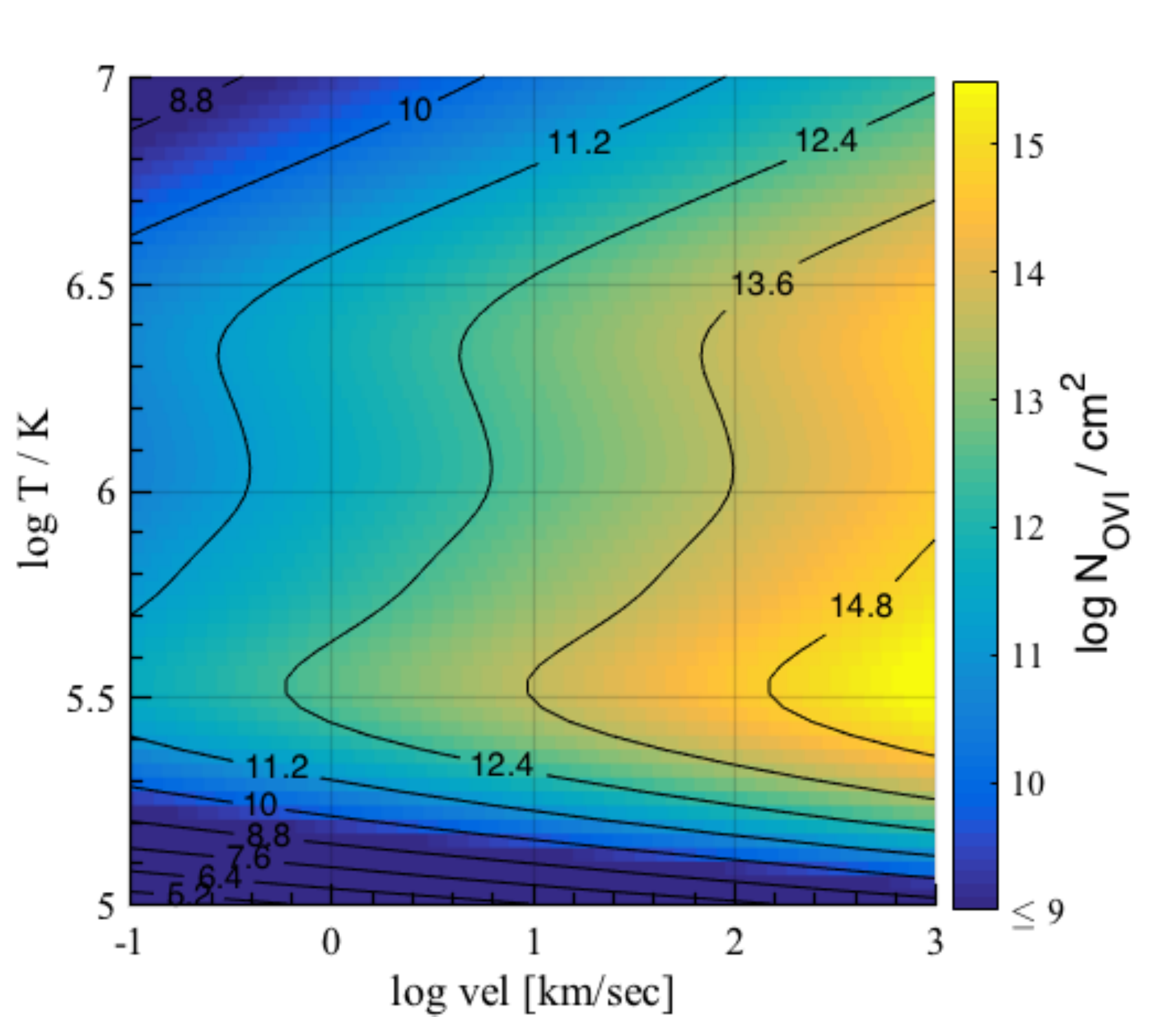}
\includegraphics[height=6.5cm,width=7.8cm]{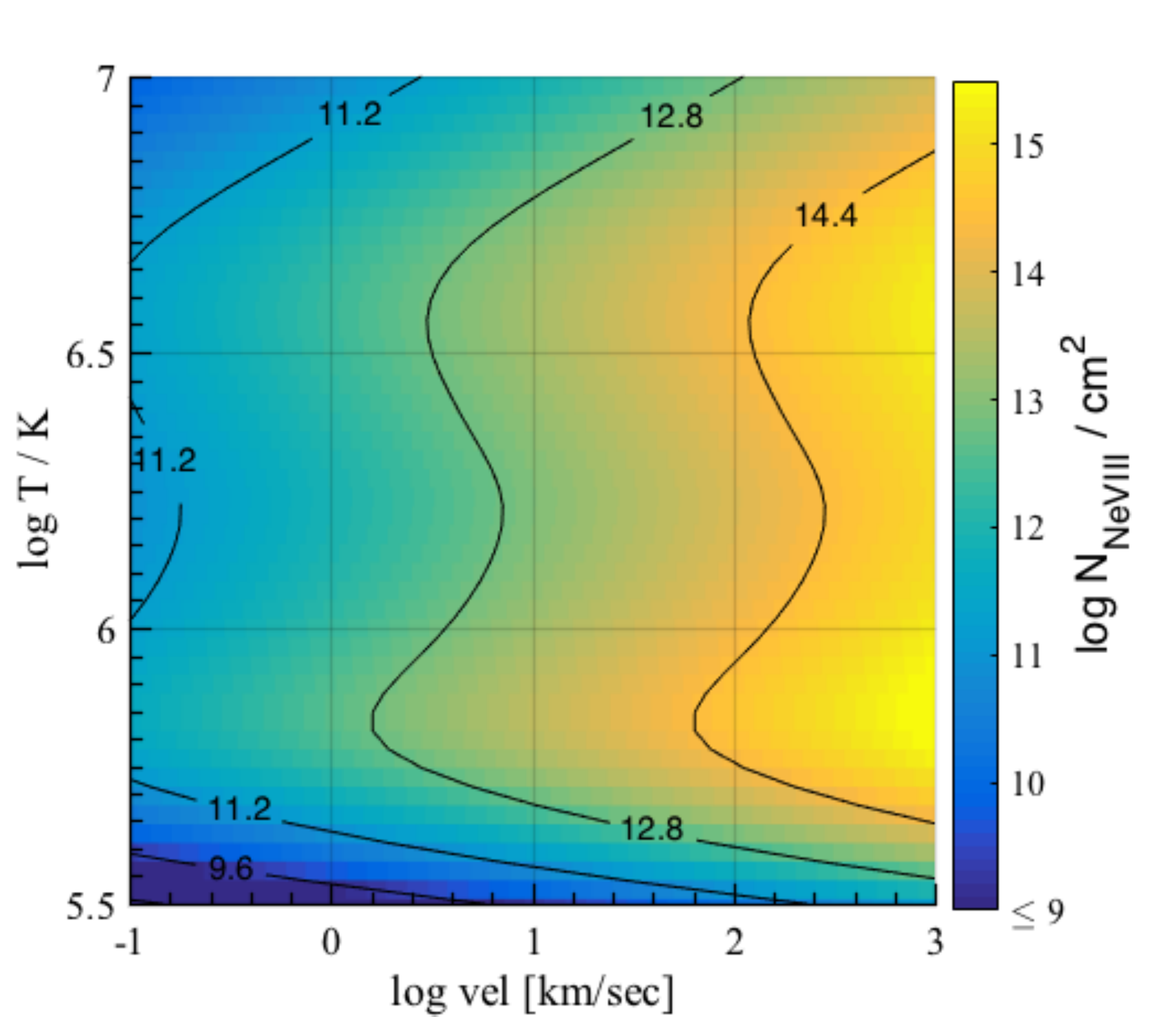}
\includegraphics[height=6.5cm,width=7.8cm]{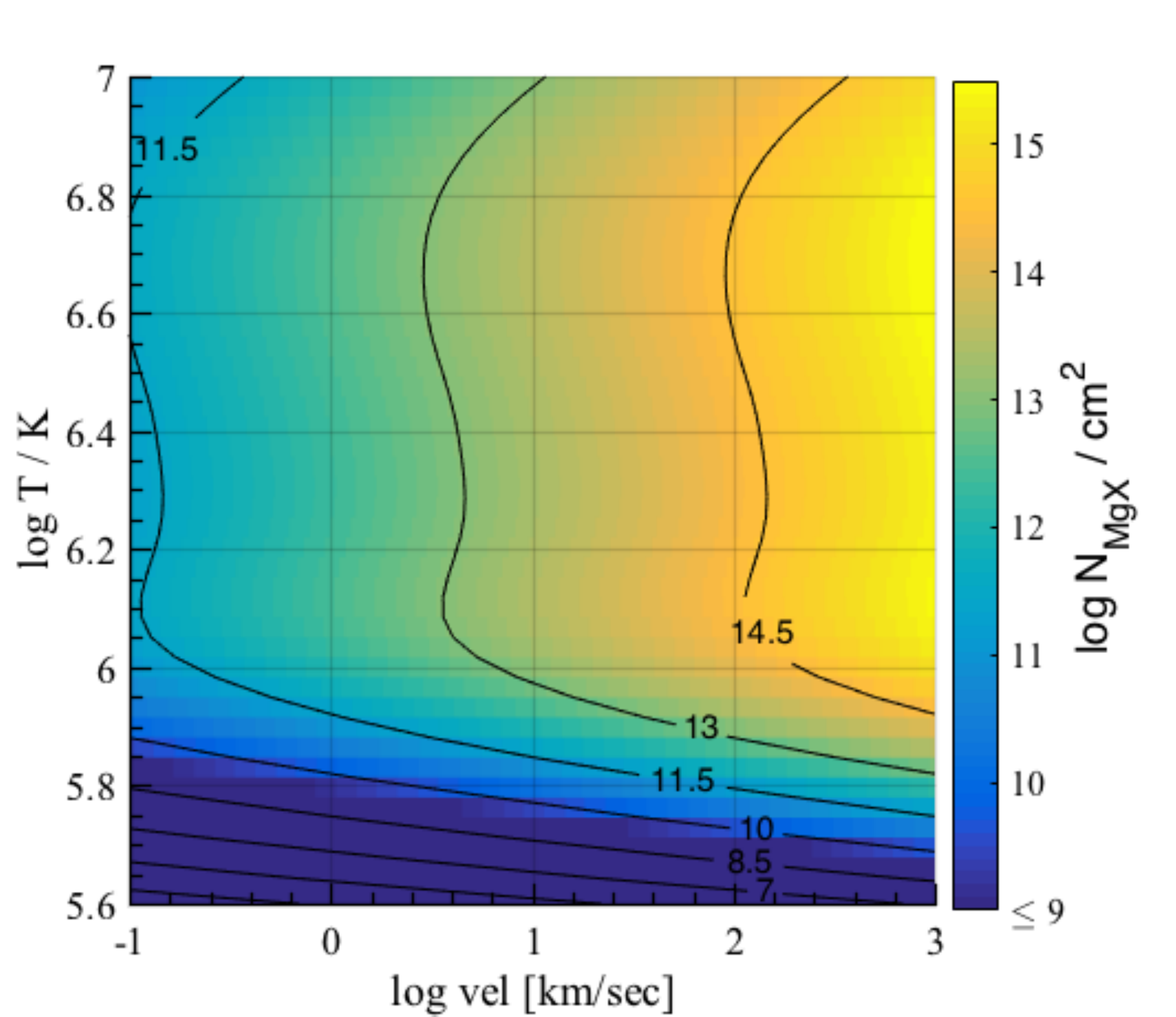}
\includegraphics[height=6.5cm,width=7.8cm]{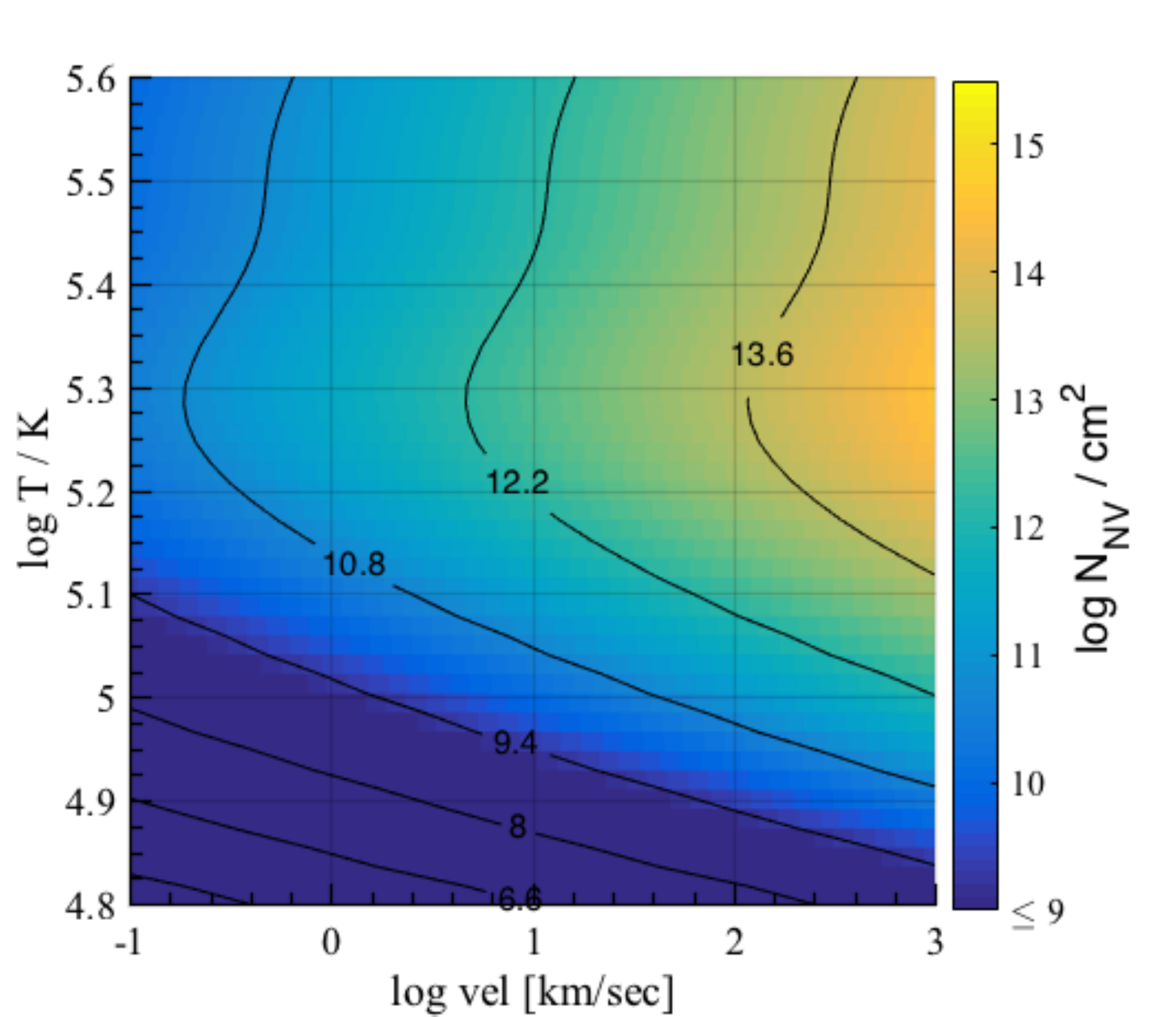}
  \caption{Cooling model prediction surfaces for {\OVI} (top left), {\NeVIII} (top right), {\MgX} (bottom left) and {\NV} (bottom right) respectively. At a fixed temperature the cooling flow column densities increase linearly with line width. At a fixed velocity the cooling flow column densities vary as a function of the fractional ionic abundance of that species. The solid black lines mark contours of constant column density.}
\label{fig:model_surf}
\end{figure*}

\section{Theory}
\subsection{The Cooling Flow Model}\label{sec:theory:coolflow}
In this section we briefly explain the cooling flow model for any radiatively cooling species that can be observed as an absorption line system. Let us consider the situation in which gas is cooling radiatively at a temperature $T$ and the number density of  ions in the gas is $n$. The total  ionic column density for that gas is given as 

\begin{equation}
N = \int \limits_{L\, O\, S} n  \;dl \;  \sim n \; L
\label{eqn1}
\end{equation} 
Here the integration is performed over the path length ($L$) of the cooling region. The total column density of that cooling gas is also written as $N_{cool}  \; = \dot{N} t_{cool}$, where $t_{cool}$ is the cooling time and $\dot{N}$ is the cooling rate per unit area. We can now write these expressions in the form $\dot{N}/n = L/t_{cool}$. We see that $\dot{N}/n$ has the dimensions of a velocity. In this paper, we are only considering those physical situations in which $\dot{N}/n$ corresponds to the actual bulk velocity of the cooling gas. We define these situations as cooling flows, and define $v_{cool} \; = \;  L_{cool} / t_{cool}\; =\; \dot{N}/n $  as the cooling velocity. We can rewrite equation \ref{eqn1} as

\begin{equation}
N_{cool} = n \; t_{cool} \;  v_{cool},
\label{eqn2}
\end{equation} 

 The cooling velocity is a characteristic speed of the cooling flow at a given temperature $T$ and it may be subsonic or supersonic with respect to the isothermal sound speed at $T$, $C_s$. In the case of a cooling instability, $v_{cool} \sim C_s$. In the case of a radiative shock, the cooling velocity is the flow velocity in the post-shock cooling region in which the cooling length $l_{cool} \sim v_{cool} t_{cool}$. In cooling galactic winds or accretion flows, $v_{cool}$ may exceed $C_s$.

Generally, in the case of radiative cooling of gas at temperature $T$ and metallicity $Z$,
\begin{equation}
t_{cool} \; =\; \frac{3 k_B T}{ n \Lambda (T,Z)} ,
\label{eqn3}
\end{equation} 
so that the product $n\,t_{cool}$ depends only on $T$ and $Z$. Here, $k_B$ is the Boltzmann constant and $\Lambda (T,Z)$ is the cooling function in the units of erg $\rm{cm^{3}\;s^{-1}}$. ( $\Lambda(T, Z)$ is often written as $\Lambda / n_{H}^2$ in the literature, e.g. \citealt{Katz1996}). We make the approximation that the electron density ($n_{e}$) is the same as the ion density ($n_{e} = n$). Using Equation \ref{eqn2}, Equation \ref{eqn3} can be rewritten as 
\begin{equation}
\boxed{
N_{cool} \; =\; \frac{3 k_B T}{\Lambda (T,Z)} v_{cool}.}
\label{eqn4}
\end{equation} 
This expression shows that $N_{cool}$ is independent of density for radiatively cooling gas.  We now move on and write down the expression for the column density of any specific radiatively cooling metal ion $X$. Taking the ionic fraction of $X$ at temperature $T_{X}$ as $f_{X} (T)$ we can write the column density of that ion $N_{X}$ as, 

\begin{equation}
\boxed{
N_{X} \; =\; \frac{3 k_B T_{X}}{\Lambda (T_{X},Z)} v_{cool} \left(\frac{X}{H}\right)_{\odot} Z\;f_{X}.}
\label{eqn5}
\end{equation} 

We know $ \left(\frac{X}{H}\right) \propto Z$. For CIE, the cooling function can be approximated as $\Lambda \; \propto \; Z^{0.7}$ over the range $\log $ T = 4.7 to 6.3 and $\log \; Z$  = -2.0 to +0.5 (Sutherland \& Dopita 1993). This corresponds to the regimes over which the primary coolants are metal lines. This then implies that $N_{X}$ only weakly depends on the metallicity. But in CIE, the ionic fraction of an ion ($f_{X}$) is typically appreciable over a narrow range of temperature centered around the temperature at which  $f_{X}$ is maximum. Therefore $N_{X}$ can be approximated by replacing $T \equiv T_{X}$ in equation \ref{eqn5}. Throughout this work, we use the collisional ionization equilibrium (CIE) cooling function computed by \cite{Oppenheimer2013}.

In equation \ref{eqn5}, $v_{cool}$ is the cooling velocity in the region where that species exist. When comparing with data, we will show in section 4 that $v_{cool}$ can be approximated by the observed line width ($\Delta v$) of the  absorption profile. We will assume that the ionized gas associated with an observed individual absorber exists at a column density weighted mean temperature ($<T>$). Using $<T>$ and the observed line width we will compute the expected model column density ($N_{X}$) from equation \ref{eqn5}. A prediction of this model is that $N_{X}$ is proportional to $v_{cool}$ (and therefore $\Delta v$), and $<T>$ is the only independent (but constrainable) parameter. Before we proceed to compare this theory to observations, we will first present some simple estimates of expected column densities for different ions with rough approximate values of $v_{cool}$ and $T$.

\subsection{Simple Estimates of Column Density}

Using equation \ref{eqn5} we can estimate the typical column density for {\OVI} gas in a cooling flow, at a temperature ($T \approx 10^{5.5}$ K) where the ionic fraction of {\OVI} is maximum ($f_{OVI} =0.22$). This yields  $N_{OVI} = 3.16 \times 10^{14}  (v_{cool} / \rm{100\, km s^{-1}} $) {\pcm}. This implies that if all {\OVI} gas exists at a constant ionic fraction ($f_{OVI}$), the observed {\OVI} column densities are directly proportional to $v_{cool}$. For example, {\OVI} created in gas radiatively cooling behind a shock, stronger {\OVI} absorption systems would be associated with higher post-shock velocities in the {\OVI} zone.

We present the cooling flow column densities for several ionized species in Figure \ref{fig:summary}, to show the relative column densities of different ionization states. To simplify these calculations, we assume that each ionization state exists at a temperature at which the ionic species has the maximum fractional abundance (in CIE).  We assume that the cooling velocity $v_{cool}$ is the isothermal speed of sound ($C_{s}$) for each species at that temperature (i.e. Mach Number, $\mathcal{M} = $1), i.e. we are assuming that the column densities of an ionization state of a gas (e.g. {\OVI}), is produced by a gas flow that is radiatively cooling from a high initial temperature to a temperature $T$ at which that ionic species has the maximum fractional abundance. We emphasize that these simple estimates are for illustrative purposes, to enable us to inter-compare the different species using the most physically natural values for $T$ and $v_{cool}$. Later, we will show that the assumption $v_{cool} \sim C_s$ is buttressed by the data for {\OVI}.

Figure \ref{fig:summary} shows the cooling flow column densities for various ionization states of Oxygen (blue squares), Neon (teal diamonds), Magnesium (red circles), Nitrogen (cyan crosses), Carbon (green diamonds) and Silicon (open diamonds), respectively. This figure presents the typical cooling column densities for each species at its peak fractional abundance temperature and shows the difference in relative column densities of different species. In such a cooling flow, the column density for {\OVI}, {\NeVIII} and {\NV} are $\log N_{\OVI} = 14.4$, $\log N_{\NeVIII} = 14.6$ and $\log N_{\NV} = 13.3$, respectively.

Equation \ref{eqn5} shows that the cooling flow column density depends on the cooling temperature and the cooling velocity of that species. Hence we can represent equation \ref{eqn5} as a 2D surface, in temperature and velocity space. Figure \ref{fig:model_surf} shows the cooling flow model predictions for four abundant species {\OVI}, {\NeVIII}, {\MgX} and {\NV}, respectively. We see that at a fixed temperature (assuming no thermal motion), the cooling column density and cooling velocities have almost a linear relationship. But at a fixed velocity, the cooling column density essentially depends on the fractional abundance of that species at CIE. At a given cooling velocity, the maximum cooling flow column density would always be observed at a temperature at which the fractional abundance of that species is maximum. The solid black lines show the contours of constant column density for each species. The enclosed numbers indicate the column density at which the contour line is drawn.

\begin{figure}
\includegraphics[height=7cm,width=9cm]{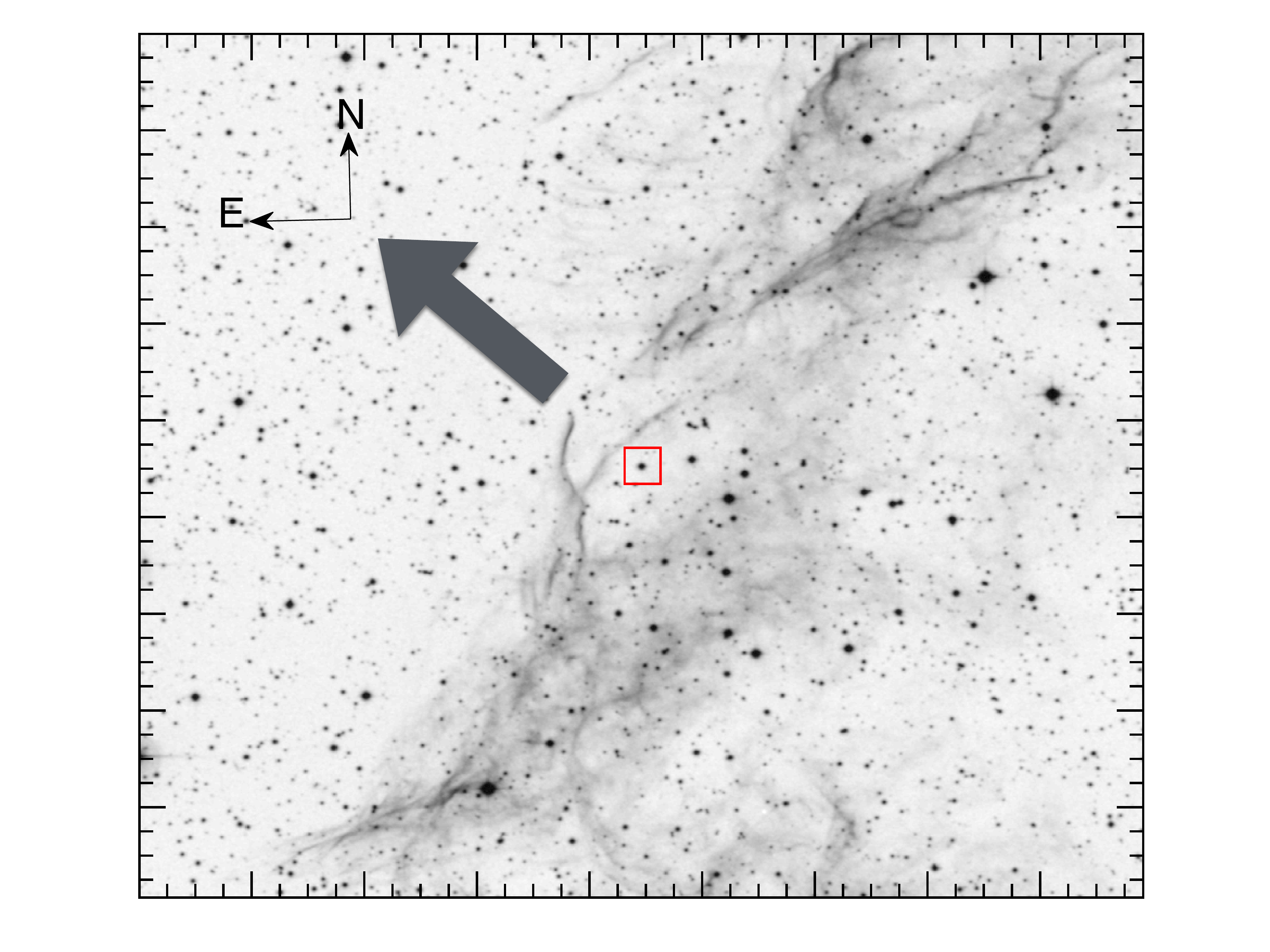}
  \caption{ The Cygnus Loop supernova remnant is shown using DSS POSS2-blue data. The filamentary structure shows the shock waves propagating outwards (direction shown by the gray arrow). In our model, warm ionized gas such as {\OVI} will be created by gas radiatively cooling behind such shock structures. The red square shows the position of blue star KPD 2055+3111, which was used to study the gas kinematics in this line of sight. The properties of the detected {\OVI} absorption-line agree well with the model of a cooling flow we describe in this paper.}
\label{fig:cygnus_loop}
\end{figure}

To better visualize the physical scenario of model presented here; in Figure \ref{fig:cygnus_loop} we show part of the Cygnus Loop supernova remnant rendered with DSS POSS2-blue data. The filamentary structures seen in the image are shock fronts propagating into the ISM, driven by a supernova explosion. As the shock wave travels outward, the gas behind the shock wave will cool radiatively and warm ionized gas such as {\OVI} will be created behind the shock. In Figure \ref{fig:cygnus_loop}, the red square marks the position of a background blue star KPD 2055+3111, which was observed with the Far Ultraviolet Spectroscopic Explorer (FUSE) spectrograph \citep{Blair2009}.  The UV spectrum was used to probe the {\OVI} gas in the Cygnus loop and \cite{Blair2009} reported that the line of sight is passing through a curved structure in the Cygnus Loop. They reported that there are two {\OVI} absorption component separated 105 {\kms} apart. The redshifted component is blended with stellar H$_{2}$ lines and therefore the only clean {\OVI} absorption that can be identified as a blueshifted component with $\rm{\log \, N_{OVI}}$ = 4.7 $\times \rm{10^{14}}$ {\pcm} and  $b_{OVI}$ = 82 {\kms} . These values are consistent with the predicted values of the model described above (see Section 5 for more discussions).

\begin{figure*}
\centering
\includegraphics[height=14cm,width=18.5cm]{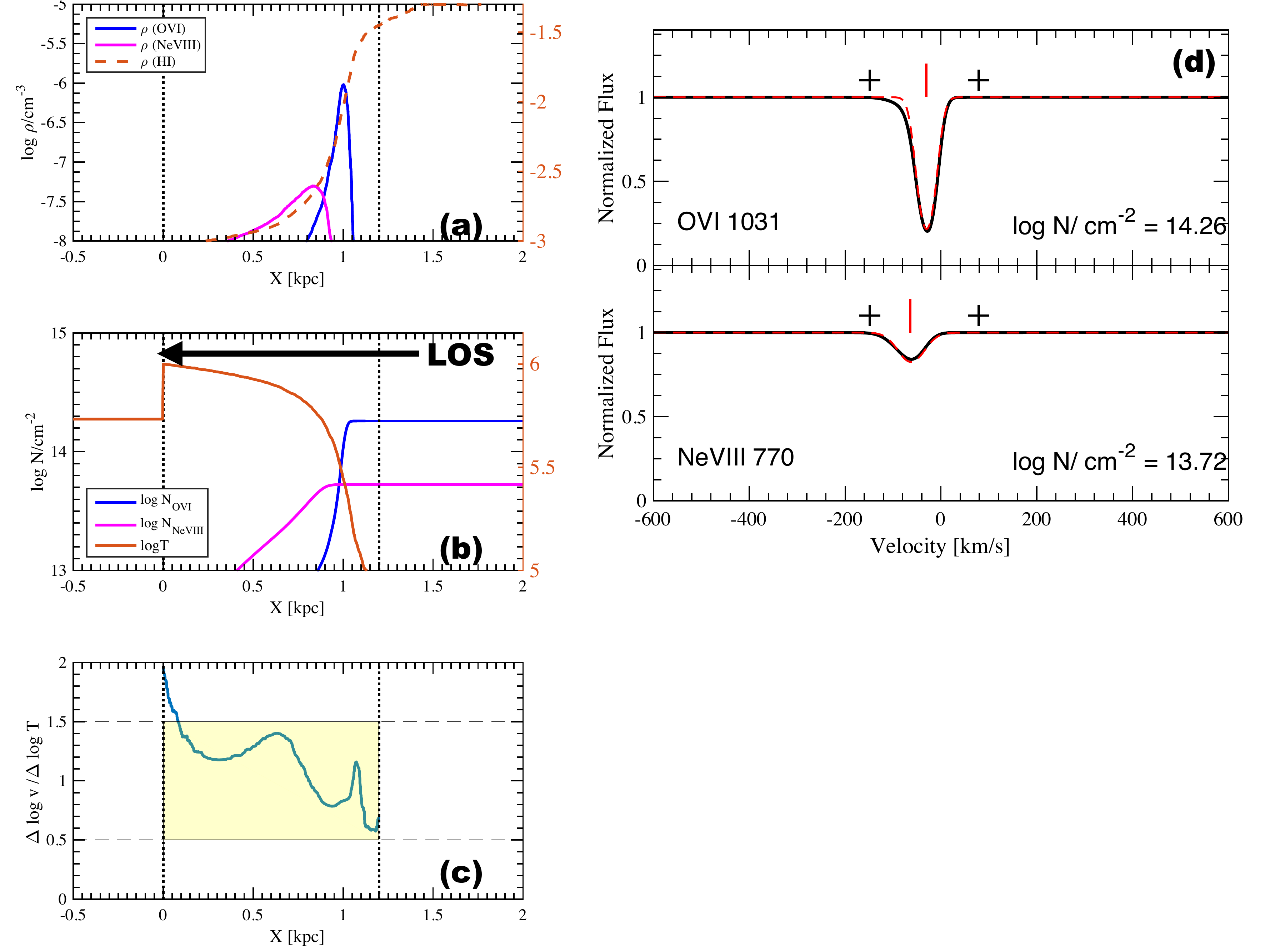}
  \caption{ 1D shock calculations showing the HI, {\OVI} and {\NeVIII} number densities (panel a), cumulative {\OVI} and {\NeVIII} column density and shock temperature profile (panel b) and $l \;=\;  \Delta \log v / \Delta \log T$ (panel c) for a shock with a shock velocity $v_{\infty}$ = 200 {\kms} and a post-shock temperature of $T \; = \; 10^{6} K.$ The yellow shaded region in panel c  shows the parameter space in the shock front where $l$ is between $l= 3/2$ and $l=1/2$. As a line of sight passes through the shock front (panel b), the corresponding {\OVI} and {\NeVIII} absorption profiles are shown in panel d. }
\label{fig:mapping_result1}
\end{figure*}

\begin{figure*}
\centering
\includegraphics[height=7.15cm,width=8.5cm]{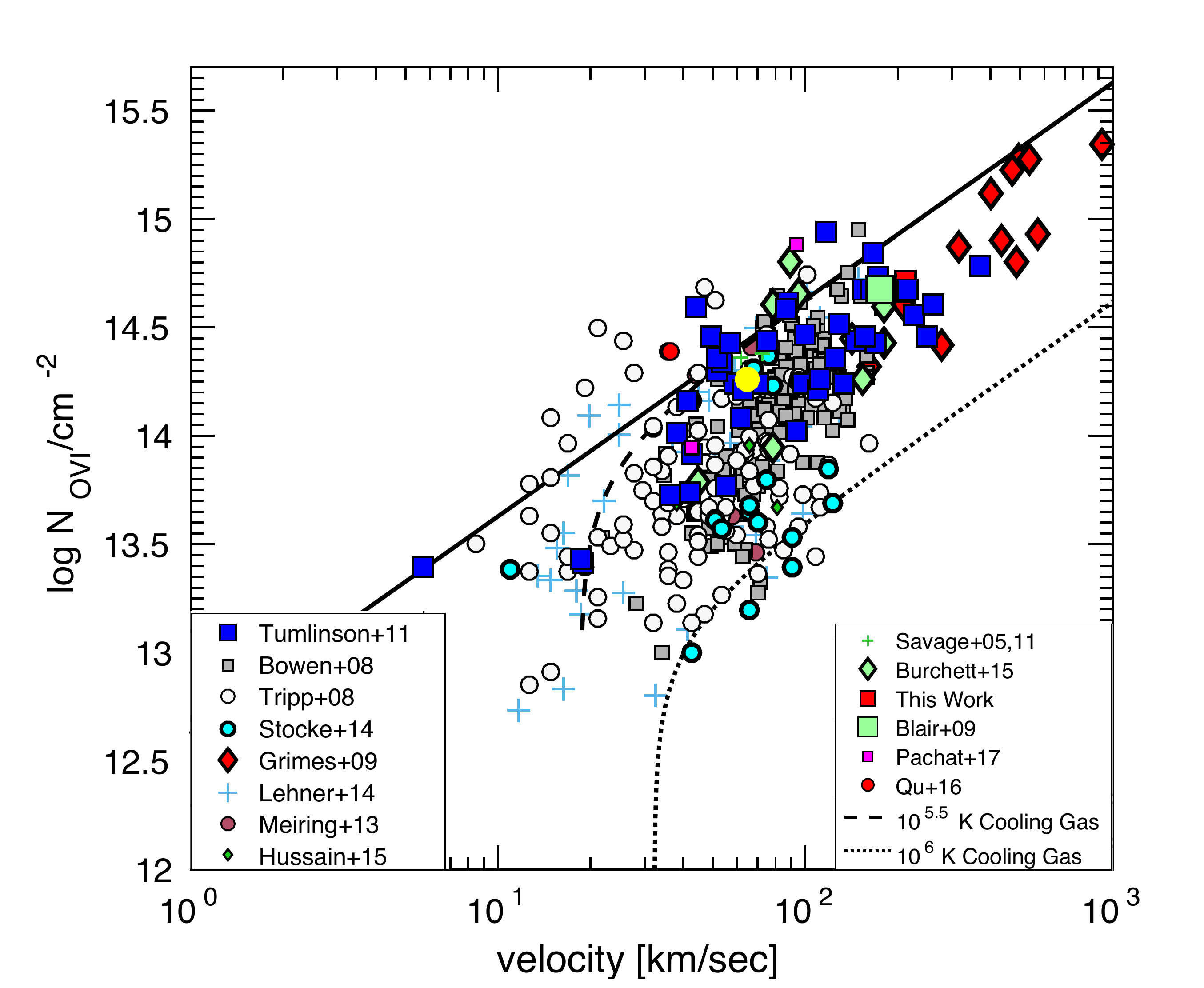}
\includegraphics[height=7.25cm,width=8.5cm]{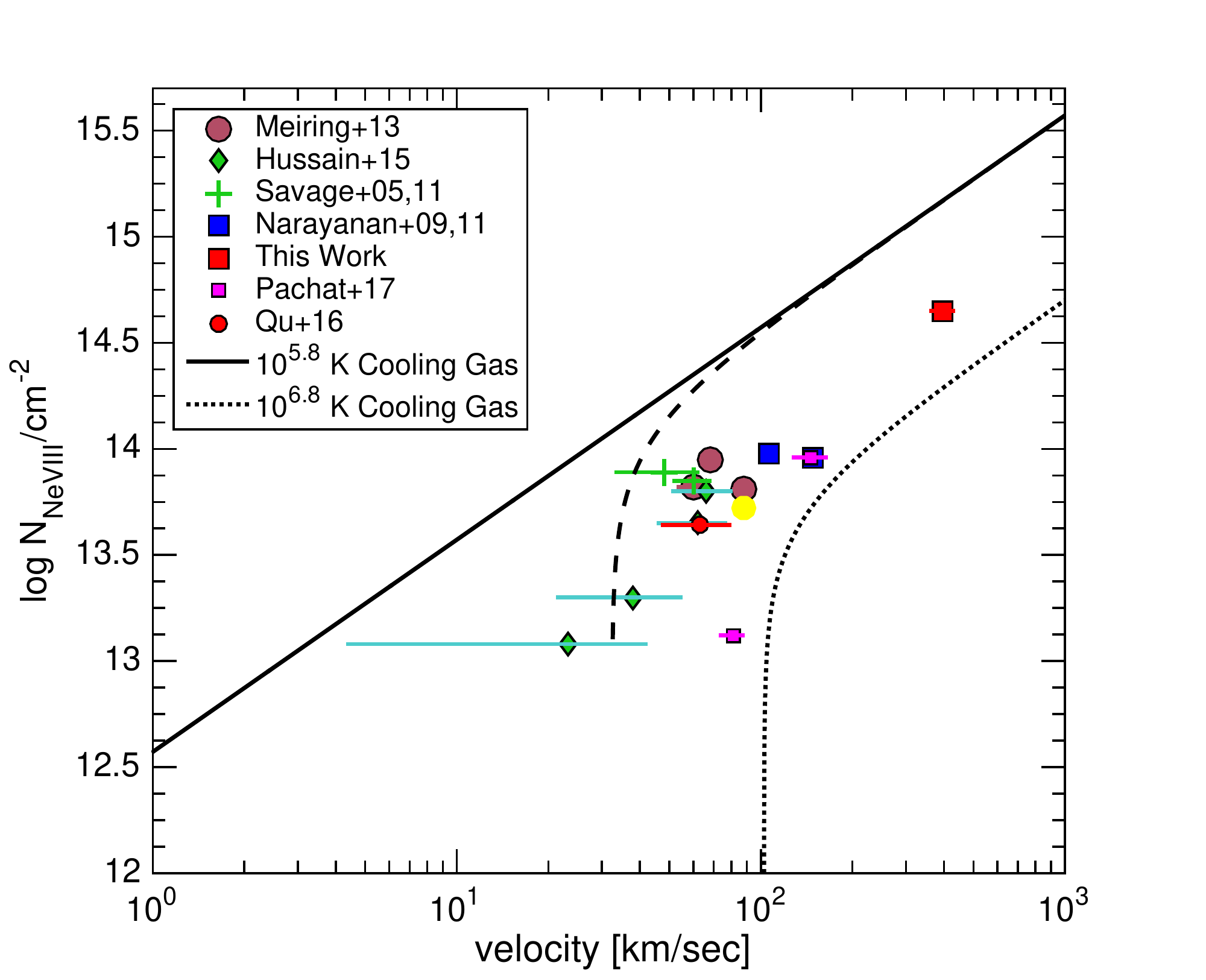}
\includegraphics[height=7.25cm,width=8.5cm]{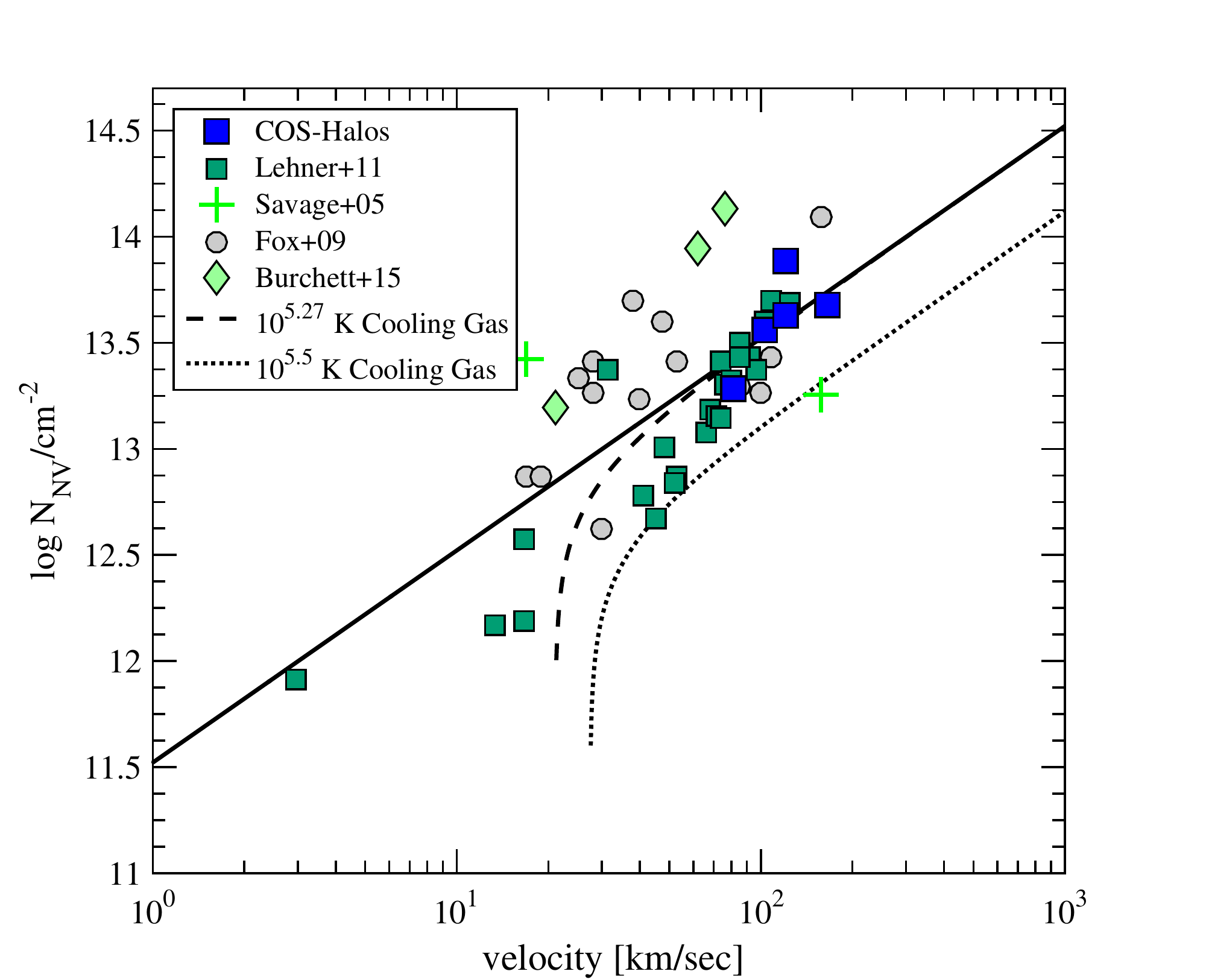}
\includegraphics[height=7.25cm,width=8.5cm]{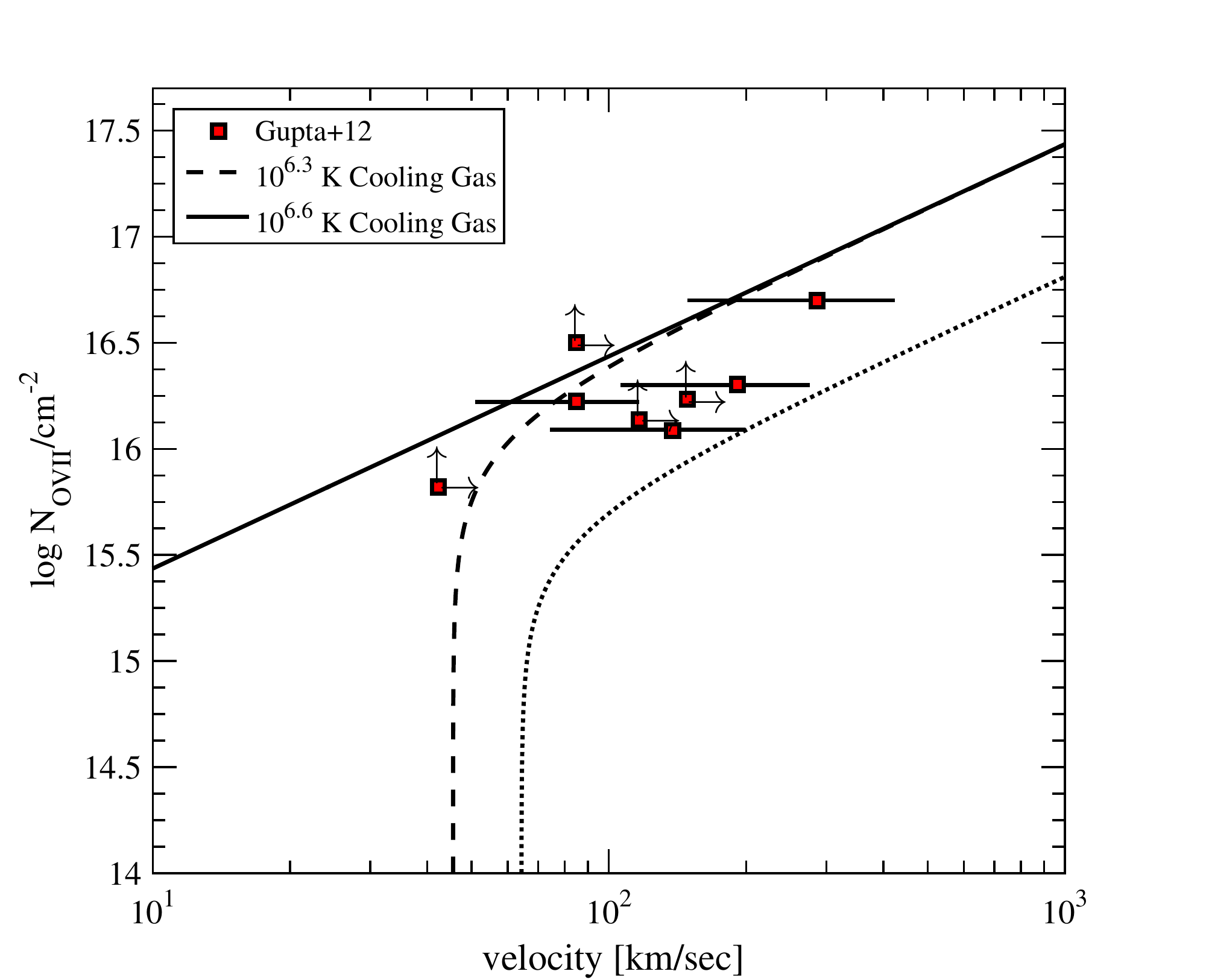}
 \caption{ Observed {\OVI} (top left panel), {\NeVIII} (top right panel), {\NV} (bottom left panel), and {\OVII} (bottom right panel)  column densities vs line widths from literature. The black lines show the predicted cooling flow column densities at different temperatures, with (dashed and dotted lines) and without (solid lines) thermal broadening, respectively. The velocity along x-axis corresponds to observed line widths for the data and $v_{cool}$ for the model predictions. The yellow circle shows the location of the O VI and Ne VIII lines produced in the numerical shock model described in section 4. The cooling flow and shock models match the observations remarkably well for a range of observations.}
\label{fig:model_compare}
\end{figure*}		

\section{Relating observed line width to cooling velocity }\label{sec:line_width}

In order to compare the model predictions with actual observations, we need to relate the observed line profile widths ($\Delta v$) to the cooling flow velocity $v_{cool}$. If there is no flow, the absorption lines have a width purely due to thermal broadening given by $v_{th} = (2kT/m)^{1/2}$ {\kms}. However for radiatively cooling gas, the observed line width is due to a combination of  pure thermal broadening and broadening due to cooling flow.  For example, in case of gas cooling radiatively behind a shock, this velocity would be the post-shock velocity in the region behind the shock, where the observed species would exist. We can write down $\Delta v$ as a combination of cooling flow velocity, and thermal broadening. 

\begin{equation}
(\Delta v)^{2} \; =\; v_{cool}^2 \;+\; v_{th}^2 \;.
\label{eqn6}
\end{equation} 

One issue that arises is how to relate the observed line width to the actual flow velocity. More precisely, the observed line width represents the spread in the flow velocity across the region where the column density of the particular ion is significant (rather than the flow velocity itself). We address this issue below where we show that along a cooling flow streamline the ratio of the line width and flow velocity is given by $\Delta v/v \sim \Delta T/T$ (the ratio of the range in temperature to the mean temperature in the region where the ionic column density is significant). If $\Delta T/T$ is of order unity, then $\Delta v/v$ is of order unity, and we can approximate $\Delta v \sim v \sim v_{cool}$, justifying the use of the line width as an estimate of the flow velocity. 

Consider a one-dimensional flow along a streamline, where the external physics drivers of the flow are changing on a sufficiently long scale length. If we assume that highly ionized species (e.g. {\OVI}, {\NeVIII}), are formed as the hot gas radiatively cools along the streamline, we can write down the conservation equations for the flow as

\begin{equation}
\dot{M} \; =\;\rho v;
\label{eqn:mass_flux}
\end{equation}

\begin{equation}
P + \dot{M} v\; =\;\Pi;
\label{eqn:momentum_flux}
\end{equation}

\begin{equation}
P  =\;\rho \mathcal{R}T.
\label{eqn:gas_law}
\end{equation} 

Here mass flux ($\dot{M}$)  and momentum flux ($\Pi$) are conserved, and $\mathcal{R} \;= (k_{B}/\mu m_{H})$. $T$ is the temperature of the cooling gas, $P$ is the pressure, $\rho$ is the density and $v$ is the velocity of the flow at temperature $T$.
Combining these equations we can write 

\begin{equation}
T  =\; \frac{v [v_{\infty} -v]}{\mathcal{R}};
\label{eqn:T_equation}
\end{equation} 
where $v_{\infty} = \Pi /\dot{M}$. The velocity $v_{\infty}$ is a fiducial velocity along the flow that has different meanings dependent on the astrophysical conditions. For example, it can be: (1) the pre-shock velocity; or (2) the terminal velocity of a cooling supersonic outflow; or (3) the terminal velocity of an accreting cooling flow. In all cases the thermal pressure at $v_{\infty}$ is assumed to be negligible. If the gas velocity in the cooling  region is significantly slowed (i.e. if $v_{\infty} \gg v$), equation \ref{eqn:T_equation} becomes
\begin{equation}
T  \sim \; \frac{v v_{\infty}}{\mathcal{R}};
\label{eqn:T_equation2}
\end{equation} 
giving a linear relationship between $T$ and $v$ (lower $T$, lower $v$);
\begin{equation}
\frac{\Delta T}{T}  \sim \; \frac{\Delta v}{v}.
\label{eqn:T_equation3}
\end{equation} 

This implies that the gas temperature in the flow changes monotonically. More exactly, using the conservation equations, $\Delta v/v = l \Delta T/T$, where $l = (v_{\infty} -  v) / (v_{\infty} - 2 v)$. Equivalently, the logarithmic derivative can be written in terms of temperature and $v_{\infty}$:

\begin{equation}
        l = \frac{\Delta\log{v}} {\Delta \log{T}} =  \pm  Y (1- Y)^{-1/2} (1 \pm (1-Y)^{1/2})^{-1}
\label{eqn:log_der}
\end{equation}

with $Y= {4 \mathcal{R} T / {v_\infty}^2}$. In the limit $v_\infty \to 0$, $|l|=1/2$, while in the limit $v_\infty \to 4 v$, as in a strong shock, $|l|=3/2$, and $|l|$ varies monotonically between these limits. Thus the relation ${\Delta v} /v  = l {\Delta T}/T $ helps justify the assumption $\Delta v \sim v \sim v_{cool}$.

In Section~\ref{sec:theory:coolflow} we noted that $v_{cool}$ may be subsonic or supersonic with respect to the sound speed in the cooling flow, $C_s$, at temperature $T$. Both cases are supported by the cooling flow model. The subsonic case is easy to understand, as the flow can be considered nearly isobaric due to sound waves crossing the cooling region. The isobaric condition together with mass conservation directly give $T \propto v$. The supersonic case is justified as follows: Combining the flow conservation equations \eqref{eqn:mass_flux} to \eqref{eqn:gas_law} yields $v_\infty = v + C_s^2 / v$. Putting $v_{cool} = v$ and multiplying by $v$, one obtains $v_{cool}^2 + C_s^2 = v_\infty v$. Comparing this to Equation~\eqref{eqn6}, we see that $(\Delta v)^2 = v_{\infty} v$. For the supersonic case, $v \sim v_\infty$, and so, here too, $v \sim \Delta v \sim v_{cool}$. We describe the interpretation of subsonic and supersonic cooling flow velocities in more detail in Section~\ref{sec:dynamics}.

In Section~\ref{sec:dynamics}, we will see that the supersonic flows seem to be mainly down-the-barrel winds and QSO sight lines through the CGM. In both cases there are geometrical arguments that could underpin the assumption that $\Delta v \sim v \sim v_{cool}$. In the case of the CGM, \cite{Heckman2017} made models of radial outflows that allow one to relate the line width to the flow speed, taking into account the variation in projection effects along a line-of-sight. In the starburst winds, we are seeing absorption against a spatially extended source (the starburst) whose size is not negligible relative to the column-density weighted size of the outflow. Again, projection effects integrated over all the lines-of-sight to the starburst will lead to a line-width $\Delta v$ of order $v$.

Looking at the fractional ionization curve for {\OVI}, we find that most of the exists in a temperature range of $\Delta T/T \; \sim $ O(1), and thus the width of the {\OVI} line in velocity would be $\Delta v/v \; \sim$ O(1). In such a case the cooling velocity at temperature $T$ is approximately the line width $\Delta v$ as defined by equation \ref{eqn6}. Thus, the cooling flow column density, $N_{cool} = n t_{cool} v_{cool}$, can be accurately estimated by using the line width, $\Delta v$, as a proxy for the cooling velocity, $v_{cool}$. 

To corroborate the above assumptions and calculations, we performed a series of time-dependent simulations of strong radiative shocks, and estimated $l$ behind the shock. For the simulations, we used the Eulerian Godunov-type (shock-capturing) hydrodynamic code PLUTO \citep{2007ApJS..170..228M}, and adapted the radiative cooling module provided within \cite{2008A&A...488..429T} to the tabulated radiative cooling function of \cite{Oppenheimer2013}. The cooling is handled by operator-splitting, whereby the rate of change of internal energy due to cooling is integrated over a hydrodynamical time step with the fourth order Cash-Karp method. Where cooling is strong, the global timestep is reduced to be at most 10\% of the cell cooling time.

The shock models are one-dimensional, and the solution is sought on a uniform grid with a spatial resolution of at least a few thousand cells per cooling length. The boundaries of the grid are set to free-flowing boundaries. We use the two-shock Riemann solver, a piecewise-parabolic reconstruction, and 3rd order Runge-Kutta time integration. The shocks are initiated with ``impulsive'' initial conditions \citep{2005A&A...438...11P}, a Riemann problem in which the upstream part of the grid is initialized with the values of the pre-shock medium, and the downstream part of the grid is initialized with the distant post-shock state of the gas, the cool dense layer at the minimum post-shock temperature.

At first, a shock wave and a rarefaction wave propagate away from the discontinuity, forming the cooling region in between. The system typically undergoes several compressional oscillations of the cooling region with periods of order the cooling time, before settling into a quasi-steady state. However, for some shock parameters the solution displays a limit-cycle behavior, the well-known cooling overstability \citep{1981ApJ...245L..23L}, whereby the cyclical compression and expansion of the cooling region remains undamped. The advantage of the impulsive shock setup is that we always remain in the frame of the mean shock location throughout the simulations and, because the distant upstream and downstream remain in momentum balance, the shock never runs off the grid.

Figure \ref{fig:mapping_result1} shows the results of this exercise. We set up a shock with a post-shock temperature = $10^6$K, density = $1.0 \times 10^{-3}$ particles/cm$^3$ and a shock speed = 200 {\kms}.  The shock Mach number is 1.8 and the shock is marginally thermally overstable. The shock structures were therefore averaged over 16 cooling times in the frame of the shock front. Panel (a) shows the HI (dashed line), {\OVI} (blue line), and {\NeVIII} (cyan line) number densities associated with the average shock structure. Panel (b) shows the cumulative column densities for {\OVI} (blue line),  {\NeVIII} (cyan line), and the shock temperature profile (red line). Panel (c) shows the calculated $l \;=\;  \Delta \log v / \Delta \log T$ values behind the shock front. The yellow shaded region in the bottom panel shows the interval in the postshock region where $l$ is between $l= 3/2$ and $l=1/2$. This is the region where equation \ref{eqn:T_equation3} is valid. Looking at the bottom and middle panels it is also clear that in this shock, most of the {\OVI} and {\NeVIII} column densities are originating from a region where $l$ is between $l= 3/2$ and $l=1/2$. The vertical dotted lines in all three panels mark the location of the shock front. Also, the post-shock regions where {\OVI} and {\NeVIII} gas are created are physically distinct and have different post-shock velocities.

We compute the synthetic {\OVI} and {\NeVIII} absorption profiles for this shock for a line of sight that is perpendicular the shock front as shown in panel (b) of Figure \ref{fig:mapping_result1}. This yields an {\OVI} absorption profile with $\log\; N_{OVI}$/{\pcm} = 14.26 and a {\NeVIII} absorption profile with $\log\; N_{NeVIII}$/{\pcm} = 13.72,  and are shown in panel (d), Figure \ref{fig:mapping_result1}. The red ticks mark the location of the line centers. If the shock moves toward the observer, the {\OVI} and {\NeVIII} absorption lines are blueshifted with respect to the undisturbed medium. The velocity gradient in the swept up gas leads to a broadening of order $ {\Delta v}/{v} \sim  {\Delta T}/{T}$. The synthetic {\OVI} line has a line width (line width containing 3$\sigma$ of the total absorption) $\Delta v$ =65 {\kms} and the synthetic {\NeVIII} line has a line width $\Delta v$ =88 {\kms}. The line ratio of {\OVI} to {\NeVIII} line is $\sim$ 0.54, similar to what is seen in observations. We plot these synthetic data points along with actual data in Figure \ref{fig:model_compare} as filled yellow circles and see that the synthetic data is similar to what we find in observations.  Because the structure of this shock is known, we will compute the {\OVI} and {\NeVIII} absorption profiles predicted by the cooling flow model. Using the mean temperatures at which bulk of {\OVI} and {\NeVIII} are created and the post shock velocities at these regions, the cooling flow model predicts the mean {\OVI} and {\NeVIII} column densities as $\log\; N_{OVI}$/{\pcm}  = 14.2 and $\log\; N_{NeVIII}$/{\pcm} = 13.75 respectively. The dashed red lines in Figure \ref{fig:mapping_result1} panel (d) show the cooling flow model predicted absorption profiles for this shock front. They are very similar to the absorption profiles computed from the numerical shock model simulation.

The fact that the line widths of {\OVI} and {\NeVIII} are significantly different is an interesting (and testable) prediction of this model. In both cases the line-widths are approximately the isothermal speed-of-sound for gas at a temperature where the ionic species reaches peak abundance ($C_{s}$ = 62  {\kms} for $T_{OVI} = 10^{5.45}$ K  {\it vs.} $C_{s}$ = 98  {\kms} for $T_{NeVIII} = 10^{5.85}$ K).

\section{Comparison with Data}
In this section we study how the predictions of the model column densities as given in equation \ref{eqn5}, compare with real data. To compare with models, we identify $\Delta v$ with a line width that contains 99.7\% (3$\sigma$) of the total absorption for any absorption line profile from its Voigt profile fit. This is described in the previous section. For any line profile quantified with a Voigt profile, and having a Doppler parameter $b_{D}$, we compute $\Delta v \;=\; 3b_{D}/\sqrt{2}$. 

Figure \ref{fig:model_compare} shows the comparison of the cooling flow column density predictions with observations for {\OVI}, {\NeVIII},  {\NV}, and {\OVII} respectively. Figure \ref{fig:model_compare}, top left panel shows the {\OVI} column densities and absorption line widths as filled points, color coded to reflect the different studies from which they are collected. To compare to the model, we need to assume a column-density-weighted mean temperature for the observed ion. We denote these by $<T>$.

\begin{figure*}[!ht]
\centering
\includegraphics[height=7cm,width=8.5cm]{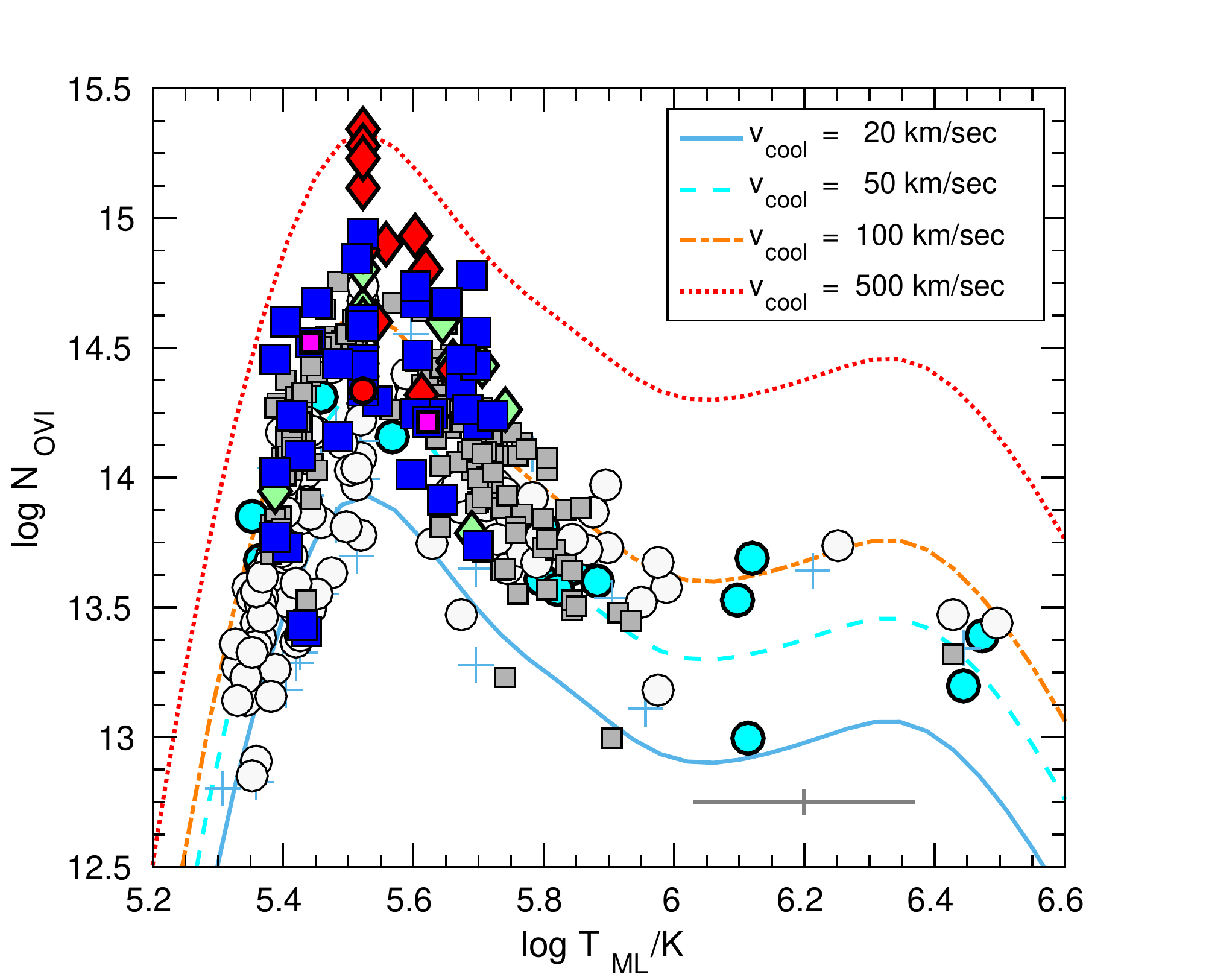}
\includegraphics[height=7cm,width=8.5cm]{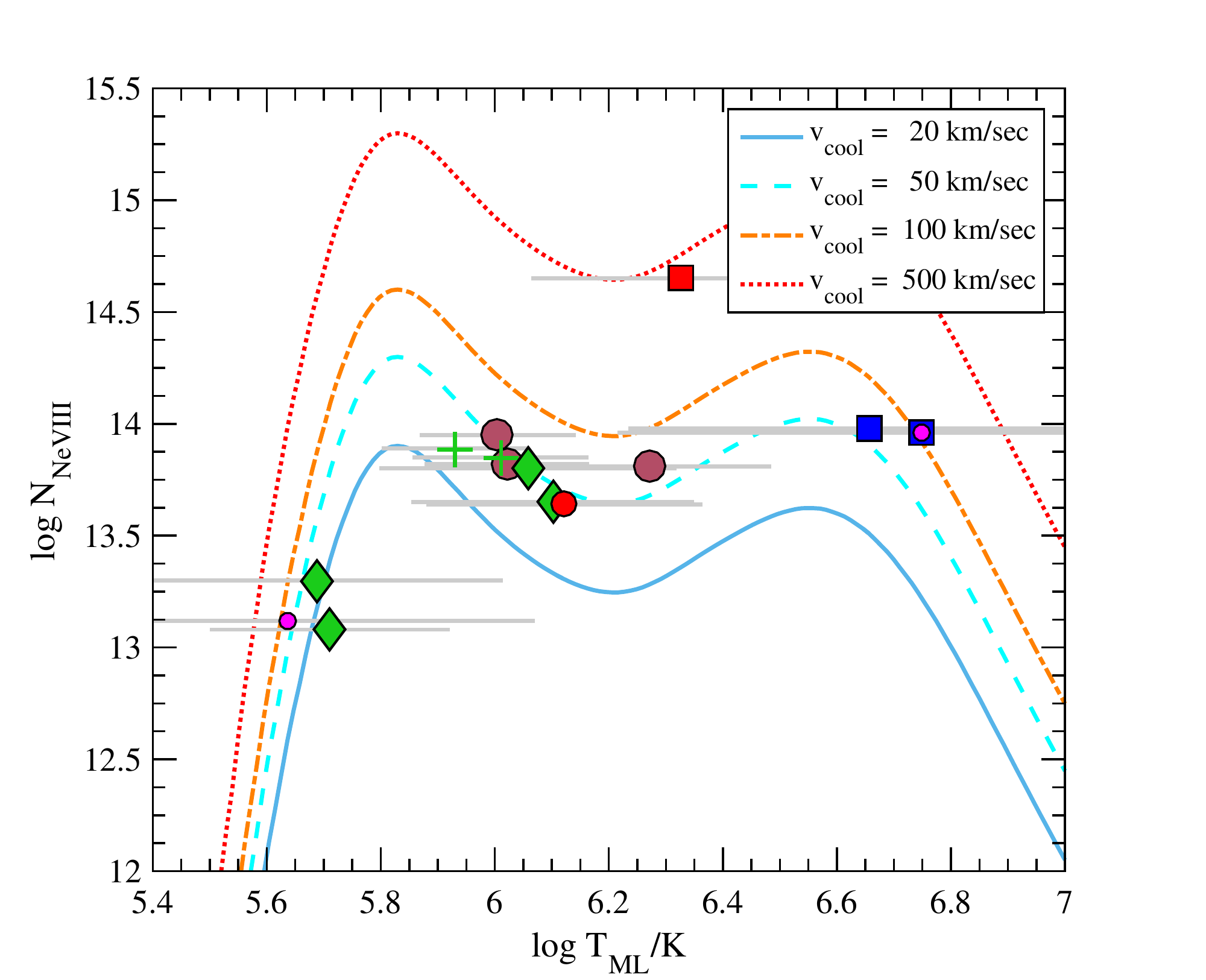}
  \caption{Maximum likelihood temperature estimates for {\OVI} and {\NeVIII} observations. All literature observations with $\Delta v >$ 10 {\kms} are included. The data points are color coded to reflect the different studies from which they are collected, as in Figure \ref{fig:model_compare}. The gray cross at the bottom of the left panel show the median $1\sigma$ uncertainty associated with $T_{ML}$ for {\OVI}. The velocity along x-axis corresponds to observed line widths for the data and $v_{cool}$ for the model predictions.}
\label{fig:temperature}
\end{figure*}	

The black lines in the top panel show the predicted cooling flow column densities for {\OVI} at $\log <T>$ = 5.5 (the temperature at which fractional abundance of {\OVI} is maximum) without thermal broadening (solid line)  and with thermal broadening (dashed line), respectively. The dotted black line shows the predicted {\OVI} cooling flow column densities at $\log <T>$ = 6.   Based on our model, the different {\OVI} absorption line systems can be divided into two regimes that spans over two decades in column densities and over one decade in line width. For absorbers with $\Delta v \gtrsim $ 30 {\kms}, the column density increases linearly with line width, while at lower line widths the column densities fall of sharply. Different {\OVI} systems preferentially occupy different part of the diagram. The {\OVI} systems of down the-barrel observations of starburst galaxies (red diamonds; \citealt{Grimes2009}), tend to have highest column densities and broader line widths.  The {\OVI} systems found around massive galaxy groups, which are possibly located in the intra-group medium (cyan circles; \citealt{Stocke14}), exhibit lower column densities and seem to be in higher temperature than the typical CGM {\OVI} absorbers (blue squares; \citealt{Tumlinson2011a}). These possible high-temperature {\OVI} systems tend to occupy the bottom middle and bottom left corner of Figure \ref{fig:model_compare}. The lack of strong ($\log$ {\OVI} $>$ 14.5) absorbers at $\Delta v \lesssim 30$ {\kms} is also a prediction of our model. Any absorber that would occupy the top left corner of Figure \ref{fig:model_compare}, could not be explained by this model. In these observations, we do not see any such systems. We also note that the properties of the {\OVI} line produced by the numerical shock model agree well with the simple analytic cooling flow model and with the data.

Figure \ref{fig:model_compare}, top right (bottom left) panels show the same comparison for {\NeVIII} ({\NV}). The black lines show the predicted cooling flow column densities for {\NeVIII} at $\log <T>$ = 5.8 and for {\NV} at $\log <T>$ = 5.27   (temperatures at which their fractional abundances are maximum) without thermal broadening (solid lines)  and with thermal broadening (dashed lines), respectively. The dotted black lines show the predicted {\NeVIII} and {\NV} cooling flow column densities at $\log <T>$ = 6.8 and $\log <T>$ = 5.5, respectively.  Both {\NeVIII} and {\NV} exhibit the correlation of increasing column density with increasing line width as seen for {\OVI} gas.

It should be noted that for the COS-Halos survey, out of 26 detected {\OVI} absorbers, only 3 lines of sight show associated {\NV} absorption \citep{Werk2013}. This should not be taken as a sign that the CGM of star-forming L* galaxies have a sub-solar nitrogen/oxygen ratio. Our model predicts that the typical {\NV} column densities should be an order of magnitude smaller than the {\OVI} column densities at similar cooling velocities (see Figure \ref{fig:summary}). In the COS-Halos survey, the typical detection threshold is $ \log \; N_{X} \gtrsim $ 13.5. Hence most of the COS-Halos {\NV} systems that are associated with {\OVI} absorption would reside below the detection threshold.  All the COS-Halos {\NV} detections are well explained by our model (Figure \ref{fig:model_compare}, bottom left panel). Moreover, higher signal to noise spectra from the \cite{Lehner2011} dataset show many more {\NV} detections in the halo of the Milky Way. The high redshift {\NV} systems detected by \cite{Fox2009} are associated with DLAs and broadly agree with our predictions, although a fraction of those {\NV} absorbers lying above the solid black curve might be photoionized systems. However, most of the {\NV} systems in Figure \ref{fig:model_compare} are consistent with the predictions from a cooling flow model.

Finally, Figure \ref{fig:model_compare}, bottom right panel shows the {\OVII} column density measurements (from X-ray spectroscopy), as a function of line widths. The arrows show the lower limits on column density and line width estimates and the range bars show the acceptable line widths from curve of growth analysis. Owing to the limits on the measurements, it is hard to quanify how well the model and the data agree, but the model predictions are qualitatively consistent with the mean trend of the data; with most of the {\OVII} absorbers falling in the linear regime of $N_{OVII}/\Delta v$ plot.

These models qualitatively reproduce the column density ranges where most of the observations lie for {\OVI}, {\NeVIII}, {\NV}, and {\OVII} respectively. The model has essentially very few free parameters apart from $<T>$ whose range of allowed values is set by simple physics.  Another feature of this model is that it naturally predicts the linear proportionality between the observed column density $N_{obs}$ and $\Delta v$ for all three species. When $\Delta v  \gg v_{th}$, the values of $N_{obs}/\Delta v$ are in the linear regime, and when  $\Delta v  \gtrsim v_{th}$, there is a turn over in the ratio of $N_{obs}/\Delta v$. This can clearly be seen in Figure \ref{fig:model_compare}.

The loci of the model predictions as a function of $<T>$ then encodes information about the values of the temperature $<T>$, flow velocity, and column density of the absorbing gas for each absorbing system. However, we know that this is an idealized situation, where we are assuming that the cooling layers are moving with the same velocity and assume that there are no small scale motions inside the layers. In reality there might be turbulent mixing layers inside the cooling flows and there might be more than one layer of cooling flow along each line of sight. If these layers are physically close to each other, they will never be seen at different velocities observationally. Moreover, as these observed absorption lines are generally not saturated, they will be linearly added together and we will infer a wrong cooling velocity if we simply use their observed line broadening parameter as a proxy of $v_{cool}$.  In addition, projection effects, non-planar flows, and other such complexities would contribute towards some of the scatter seen in Figure \ref{fig:model_compare}. For simplicity we will refrain from modeling these effects in this paper.

A useful application of this model is to estimate the temperature at which bulk of the highly ionized gas would be created. For an absorber the likelihood of observing $N_{obs}$ and line width $\Delta v$ given the model can be written as

\begin{equation}
\begin{split}
\mathcal{L}(N_{X},v_{cool})\; \equiv  \frac{1}{4 \pi^{2} \sigma^{2}_{N}\sigma^{2}_{v}} & \exp\left[\frac{-(N_{obs}-N_{X})^2}{2 \sigma_{N}^{2}}\right]   \\
    &  \exp\left[\frac{-(\Delta v-v_{cool})^2}{2 \sigma_{v}^{2}}\right]
\end{split}
\end{equation} 

Here $N_{X}$ is the cooling flow model column density given by equation \ref{eqn5} and $v_{cool}$ is the cooling velocity (after accounting for thermal broadening) given by equation \ref{eqn6}. $\sigma_{N}$ and $\sigma_{v}$ are measured uncertainties on  $N_{obs}$ and $\Delta v$ respectively.  To compute the maximum likelihood temperature ($T_{ML}$) at which bulk of the highly ionized gas is created we marginalize over $v_{cool}$ and compute the temperature at which the marginalized $\mathcal{L}$ is maximum ($\mathcal{L} (T_{X}) = \int \;  L(N_{X}(T_{X},v_{cool}),v_{cool}) \; dv_{cool} $).

Figure \ref{fig:temperature} shows the maximum likelihood temperature estimates ($T_{ML}$) for {\OVI} and {\NeVIII} absorption lines collected from the literature. The data points mark the estimated $T_{ML}$ for each individual observation, and the colored lines show the predicted model column densities at that given temperature for a fixed cooling velocity.  The data points are color coded to reflect the different observations from which they are obtained, as in Figure \ref{fig:model_compare}.  For {\OVI} most of the measurements cluster around $\log T_{ML}\; \approx$ 5.5, which is also the temperature at which the {\OVI} fractional abundance is maximum.  The gas temperatures of {\OVI} absorbers detected in CGM of $L*$ galaxies (blue squares, from the COS-Halos survey), are estimated to be within $5.4 \leq \log T_{ML} <5.8$.  The cross in the bottom right corner of Figure \ref{fig:temperature} left panel shows the median $1\sigma$ uncertainty associated with the $T_{ML}$ estimates. We compute the $1 \sigma$ uncertainty for each $T_{ML}$ by measuring the temperature range in the marginalized likelihood function which contain 68$\%$ of the area of its likelihood function around $T_{ML}$. The median $1\sigma$ uncertainty for {\OVI} $T_{ML}$ estimates is $\sim$ 0.17 dex. Curiously, we also find a small subsample of {\OVI} absorbers at $\log T_{ML}\; >$ 6 with lower column densities. The blue cyan circles with $\log T_{ML}\; >$ 6 are {\OVI} measurements from \cite{Stocke14}, who found that these absorbers are primarily detected around galaxy groups. The open circles are from the absorption selected {\OVI} survey of \cite{Tripp2008} and we don't have any information on whether they are located in the CGM or IGM.  In all cases we find that {\OVI} systems associated with $\log T_{ML}\; >$ 6, have lower column densities compared to typical {\OVI} absorbers seen around star-forming $L*$ galaxies (e.g. COS-Halos). Our model predicts that these absorbers are at higher temperatures as they have low {\OVI} column densities for their observed line widths.  For such intra-group gas, it is plausible that the gas might be prevented from cooling further and the stronger {\OVI} absorption phase (seen at 10$^{5.5}$ K) will not arise.  Alternatively it is also possible that we simply happen to observe only a statistically small subsample of {\OVI} absorbers, that have not yet cooled to a low enough temperature to create 10$^{5.5}$ K {\OVI} gas.  And in future, as the gas cools higher column density 10$^{5.5}$  {\OVI} systems would also be created there.

Figure \ref{fig:temperature}, right panel shows the $T_{ML}$ for the small number of  {\NeVIII} observations found in the literature. We find a more scattered distribution of $T_{ML}$, with most of the absorption lines being consistent with having $\log T_{ML} \; >$ 5.8. As the number of absorbers are small, we show individual $1\sigma$ uncertainties on each $T_{ML}$ estimate.

For {\NV} typical $T_{ML}$ estimates are between $5.1 < \log T_{ML}\; <5.6$. However, the median uncertainties on $T_{ML}$ are quite large. We find that typical $1\sigma$ uncertainty on {\NV} $T_{ML}$ is $\gtrsim$ 0.5 dex.

\begin{figure}
\includegraphics[height=8cm,width=10cm]{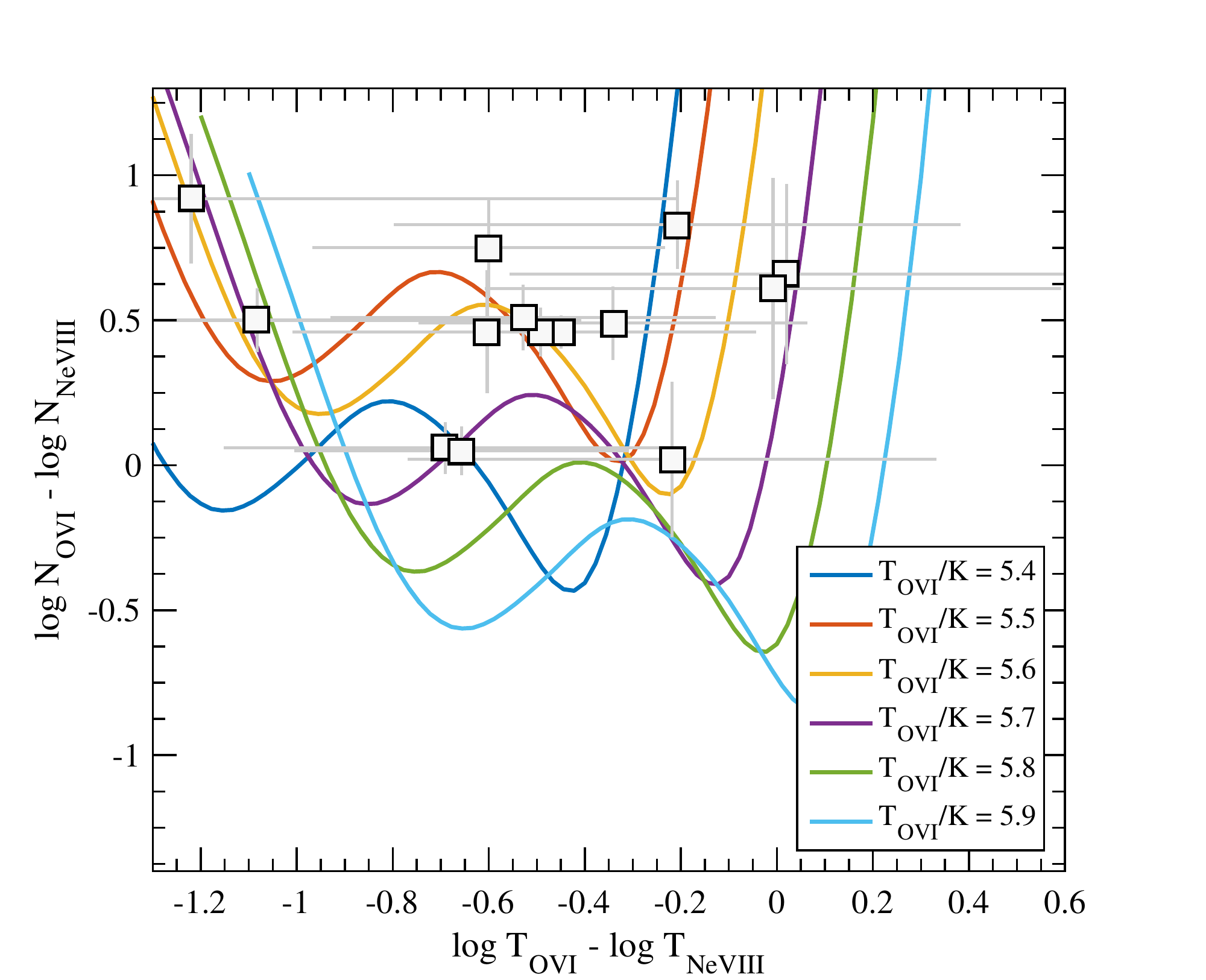}
  \caption{The line ratios of observed {\OVI} and {\NeVIII} absorbers at their respective maximum likelihood ($T_{ML}$) temperatures. Each colored curve show the line ratios predicted by this model (for a given {\OVI} temperature) plotted as a function of the difference in temperature between {\OVI} and {\NeVIII}. The observations are consistent with the model predictions. In particular, the temperatures estimated for {\NeVIII} are generally higher than for {\OVI}. }
\label{fig:line_ratio}
\end{figure}

We proceed to use the estimated $T_{ML}$ for these ions and compare the column density ratios of a subset of 10 absorbers for which both {\OVI} and {\NeVIII} absorption lines were observed. The white squares in Figure \ref{fig:line_ratio} show the observed column density ratios of {\OVI} and {\NeVIII} absorbers and the difference between their respective maximum likelihood temperatures. The error bars show the $1\sigma$ uncertainty associated with the measurements. Most of the observed {\OVI} absorbers have slightly higher column densities as compared to {\NeVIII} absorbers. Also {\NeVIII} absorbers are at a higher $T_{ML}$ as compared to {\OVI}. The solid colored lines are model predictions for {\OVI} and {\NeVIII} absorption column densities at different temperatures. Different colored solid lines indicate the temperature at which {\OVI} gas was created. We find that the model predictions are consistent with the observed column density ratios for these absorbers. 

We conclude that this method allows us to estimate the temperature of the absorbing gas in an independent manner. It can also be applied to many other species and we can predict the typical cooling flow column densities that could be observed as a function of the temperature of the cooling gas. In Figure \ref{fig:predictions} (Appendix B) we summarize our predictions for O IV, V, VII, C V, Ne V, VI, Mg X, and Si XII, respectively. Different lines in each panel represent the predicted column densities for different cooling velocities. The different velocity curves are assuming no thermal motion for this gas. All species show a maximum cooling flow column density at a temperature where it has maximum fractional abundance. Some of the species can be observed with HST/COS UV spectroscopy, whereas some others can be accessed by X-Ray observations. Particularly for species like Mg X and O IV, we would be able to start comparing these model predictions with observations in the coming years, when new UV absorption line measurements will be published.

\begin{figure}
\centering
\includegraphics[height=8cm,width=10cm]{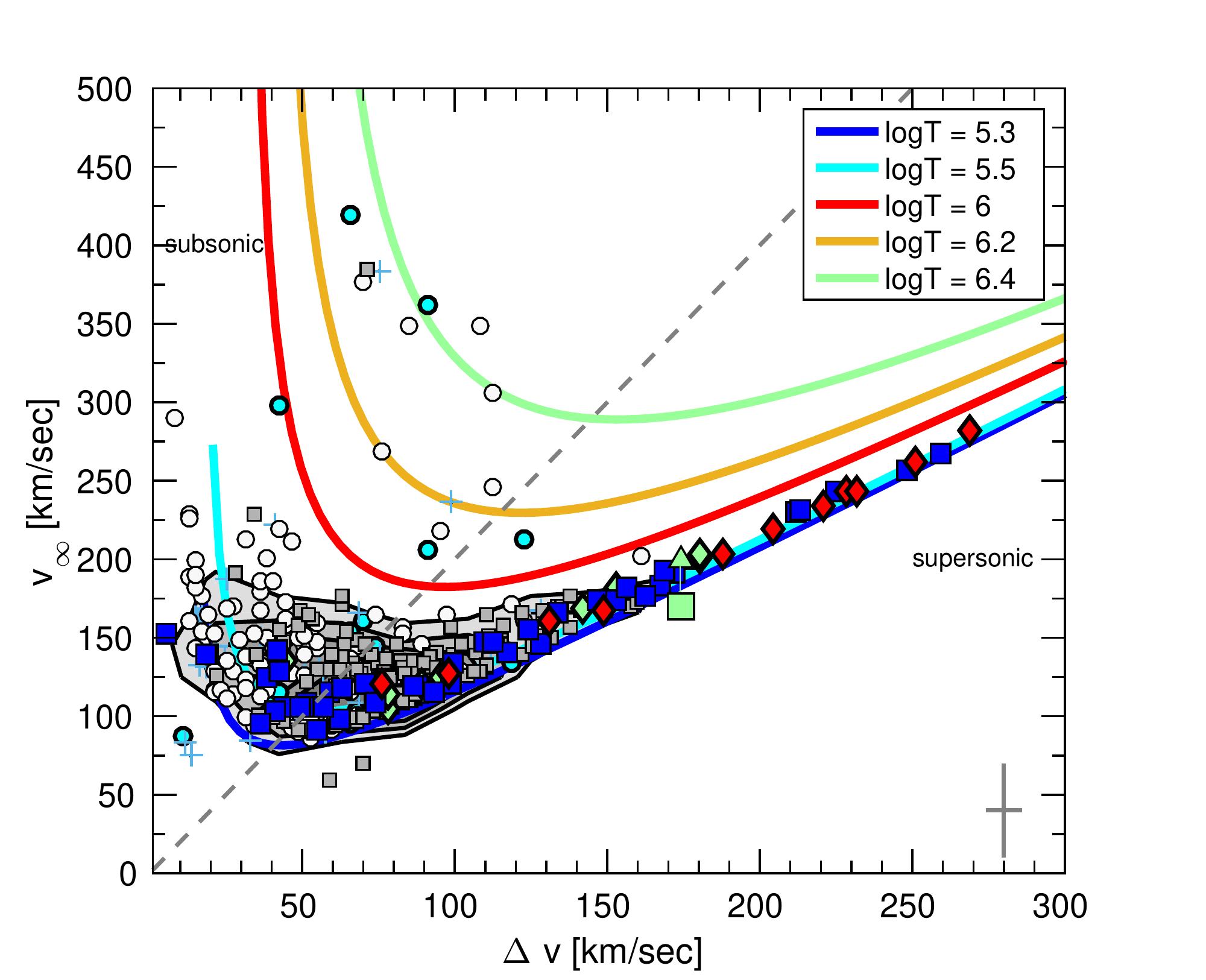}
  \caption{Estimated values for $v_\infty$ for {\OVI} measurements. The measurements are color coded as in Figure \ref{fig:model_compare}. The underlying contours show the distribution of the values for $v_\infty$  for the {\OVI} measurements.  The gray cross at the bottom right corner shows the median uncertainties associated with estimated values for $v_\infty$ and the observed $\Delta v$. The isotherms shown as solid lines are the estimated values for $v_\infty$ at different cooling temperatures. The isotherms are given by Equation~\eqref{eqn:shock_vel} and are loci of monotonically increasing Mach number $\mathcal{M} = v_{cool} / C_s$ from left to right. The diagonal dashed line is the sonic line ($\mathcal{M} = $1). Points to the left of the curve minima are subsonic cooling flows, whereas the points to the right are supersonic cooling flows. We identify the subsonic region with radiative shocks, where $v_\infty$ represents their shock speed, and the supersonic region with cooling winds or accretion flows, where $v_\infty$ represents their terminal velocity.}
\label{v_shock_delta_v}
\end{figure}	


\begin{figure}
\includegraphics[height=8cm,width=10cm]{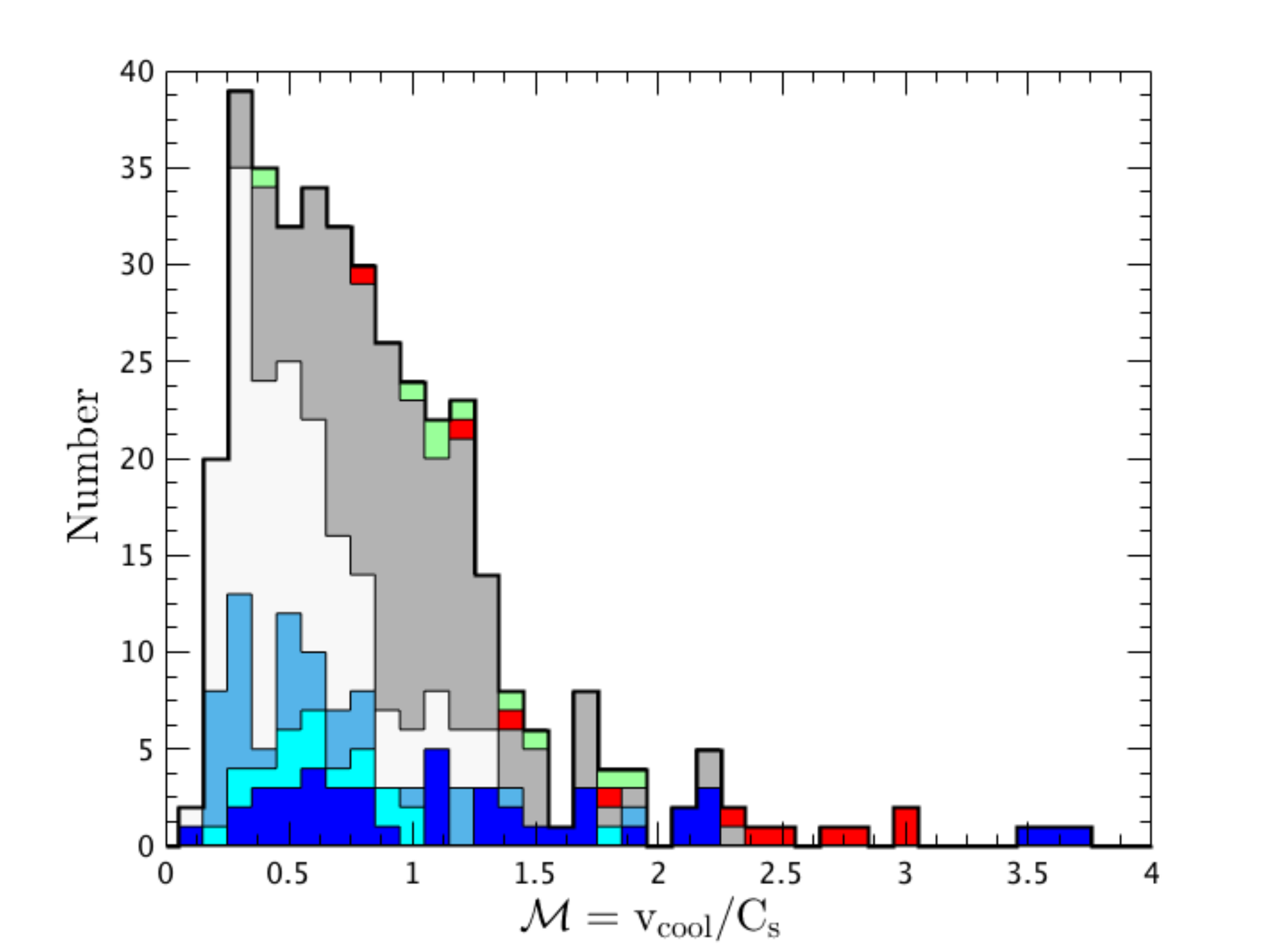}
  \caption{Histogram showing the Mach numbers ($v_{cool} / C_s$) of the flow associated with each {\OVI} observations with a temperature $T$. The bulk of the {\OVI} absorbers are associated with a subsonic flow with a median Mach number of 0.7. The small fraction of {\OVI} absorbers associated with supersonic flows are mostly arising in starburst winds seen down-the-barrel. The histogram is color coded as in Figure \ref{v_shock_delta_v}.}
\label{fig:mach_number}
\end{figure}	


\section{The dynamics of the cooling gas}\label{sec:dynamics}
One application of this model is to independently constrain the dynamics of the gas in the cooling flow. In particular, we can differentiate between the cases where $v_{cool}$ is subsonic and supersonic.  The relevant conservation equations are given by equations \ref{eqn:mass_flux}, \ref{eqn:momentum_flux} and \ref{eqn:gas_law}. These equations can be combined to yield
\begin{equation}
\frac{\Pi}{\dot{M}}  =\; \frac{ \mathcal{R}T}{v} + v = v_{\infty}.
\label{eqn:shock_vel}
\end{equation} 
Thus, for any species, knowing the temperature and the cooling velocity, we can estimate $v_\infty$ associated with that cooling flow. Recall that, depending on the astrophysical conditions, $v_\infty$ may be the preshock velocity of a radiative shock, or the terminal of a cooling wind or accretion flow. Figure \ref{v_shock_delta_v} shows the predictions for $v_\infty$ according to Equation~\eqref{eqn:shock_vel} for cooling {\OVI} gas at temperatures ranging from $\log T = 5.3$ to $\log T = 6.4$. To compare with data, we are showing $\Delta v$ along the x-axis, as defined by equation \ref{eqn6},  keeping in mind that $v \sim \Delta v \sim v_{cool}$. The estimated values for $v_\infty$ from the {\OVI} absorption line observations are shown as colored symbols. We can estimate the values for $v_\infty$ for these observations as we have computed the maximum likelihood temperatures for these species in section 5. The data points are color coded as in Figure \ref{fig:model_compare} to identify the source of these measurements. The underlying contours show the distribution of the {\OVI} velocities as compared to the model predictions. The CGM {\OVI} absorbers from COS-Halos (dark blue squares) and starburst galaxies (red diamonds), are in the right part of the diagram, where they follow a linear relation with $v_{\infty}$. 
To gain more insight, equation \ref{eqn:shock_vel} can be rewritten as
\begin{equation}
v_{\infty} = v[1/(\mathcal{M}^{2}) + 1]
\label{eqn:mach}
\end{equation}
Here, $\mathcal{M}$ is the Mach number ($v_{cool} / C_s$) of the cooling flow at the location with temperature $T$. We emphasize that this is the Mach number of the cooling flow, not the Mach number of a shock or the Mach number given by the terminal speed of a wind. If $\mathcal{M} < 1$, the cooling flow is the downstream flow of a radiative shock. If $\mathcal{M} > 1$, the cooling flow is a cooling galactic winds or accretion flow. The isotherms in Figure \ref{v_shock_delta_v} then represent loci along which the Mach number increases from the subsonic (upper left) to trans-sonic (near the curves minima) to supersonic (far right) regimes. In Figure \ref{fig:mach_number}, we show a histogram of $\mathcal{M}$. We see that the vast majority of cases are sub-sonic or trans-sonic. These would correspond to gas cooling through the {\OVI} regime behind a shock, and populate the top left branch in Figure \ref{v_shock_delta_v}. For these absorbers, the quantity $v_\infty$ is the shock velocity, and the cooling flow is significantly slowed from the pre-shock velocity ($v_{\infty} \gg v$). The typical shock velocities in these systems imply post-shock temperatures of $T_{shock} \sim 10^{5.2-5.8}$ K, high enough to produce {\OVI}. The {\OVI} absorbers along the upper right branch are in supersonic cooling flows, mostly arising in starburst winds. For these absorbers, $v_\infty$ corresponds to the terminal velocity of the wind. They are also associated with similar temperatures as the post-shock temperatures of the subsonic cases. It is important to note, however, that there may be bulk subsonic cooling flows and subsonic galactic winds on the top left branch, as well as some post shock regions in which the cooling of the OVI occurs where the flow is still supersonic with respect to the {\OVI} sound speed. The latter arises naturally in thermally overstable shocks (as we have also found in our simulations), in which the cooling in the downstream flow is sufficiently rapid to induce velocity drops that generate secondary shocks in the cooling layer \citep{Falle1975}.

Observationally, the shock velocity of the Cygnus loop is found to be $\gtrsim$ 170 {\kms} \citep{Danforth2001}. This is shown as the green square in Figure \ref{v_shock_delta_v} and is consistent with the model predictions shown as the solid lines. 

A remarkable feature of Figure \ref{v_shock_delta_v} are the minority of systems on the hot branch at $ \log T_{ML} \;> $6. These are the higher temperature lines found in Lithium-like systems due to the {\OVI} populated by recombination from the {\OVII} state. Generally these are from hotter and therefore more massive systems. The total mass inferred from these systems would be much larger and would be dominated by the hotter {\OVII} phase since the fraction of oxygen in the {\OVI} state is low relative to the cooler {\OVI} systems. The open and cyan circles that constitute these hotter systems are found to be associated with the IGM and groups rather than with the general CGM \citep{Stocke14}. This then is a consistent physical picture.

\section{Conclusions}
Diffuse gas residing in a warm/hot ($T \sim 10^{5-6}$ K) phase is important in many astrophysical contexts, but until recently it has been extremely difficult to observe.  With the introduction of the HST/COS ultraviolet spectrograph, we now have a large enough sample of absorbers to attempt a unified analysis of all such observations.  In this paper, we showed that {\OVI}, {\NeVIII}, {\NV} and, {\OVII} absorption line systems, observed in diverse environments such as the disk of the Milky Way, Milky Way halo, starburst galaxies, the circumgalactic medium and the intra-group medium at low and high redshifts, can be explained by a simple model where gas radiatively cools in a flow in collisional ionization equilibrium with a cooling velocity, $v_{cool}$.

We collected a representative sample of {\OVI}, {\NeVIII}, {\NV}, and {\OVII} absorption line systems from the literature.  Based on our model, the different {\OVI} absorbers can be divided into two regimes that span over two decades in column density. The observed {\OVI} column density linearly increases with line width for broad ($\Delta v \gtrsim 30$ {\kms}) lines and turns over and steepens for narrower lines.

The cooling flow model presented in this work can qualitatively reproduce these observed column densities at their respective line widths. We associated the line width with the velocity of the flow at the point where the line is cooling in, e.g., OVI, and related this to the column density of the ion. The column densities predicted by this model are independent of metallicity for any system that has $[\mathrm{O}/\mathrm{H}] > 0.1 [\mathrm{O}/\mathrm{H}]_{\odot}$. For any given highly ionized absorber at an observed column density and line width, we compute a maximum likelihood temperature associated with that absorber. 

We were able to simultaneously explain absorption line systems observed in diverse environments, over a range of temperature, and covering ions ranging in ionization potential from {\NV} to {\OVII}. This model naturally explained the ubiquitous presence of {\OVI} around $z \sim 0.2$ star-forming L* galaxies, and the surprisingly small strength of {\NV} around the same set of galaxies in the COS-Halos survey. We predict that on average {\OVI} should have an order of magnitude higher column density than {\NV}. The detection threshold for the COS-Halos survey is such that most of the associated {\NV} could not be detected due to lower S/N spectra. Along lines of sight with higher S/N spectra, probing the halo of the Milky Way, such {\NV} systems are detected with higher frequency and are consistent with our model predictions.

This model successfully reproduces the observed column density ratios of {\OVI} and {\NeVIII} absorbers.  Most of the observed {\OVI} absorbers have slightly higher column densities relative to the observed {\NeVIII} absorbers. Further, {\NeVIII} absorbers are created at higher maximum likelihood temperatures as compared to the {\OVI} absorbers.

We also predict column densities that can be seen with other ions such as OIV, V, VII, CV, NeV, VI, {\MgX}, and {\SiXII}. The small number of {\OVII} observations are consistent with our model predictions. Future observations using HST/COS spectroscopy for the UV lines, or X-Ray absorption observations of the highly ionized hot gas will allow us to constrain this model further and test the origin of such absorption line systems.

Using the conservation equations for the cooling flow, we studied the detailed physics of the line formation in CIE as a function of temperature and showed that the temperature change across the line ($\Delta T / T)$ is fixed and of order O(1). At the same time, we have shown that the cooling velocity $v_{cool}$, which contributes to the line width $\Delta v$, is a good approximation to the flow velocity in the cooling region, and that this is valid for both subsonic and supersonic cooling flows. These approximations were confirmed with hydrodynamic simulations of radiative shocks. It is this estimate for $v_{cool}$ through the line width that is robustly direct from the observations that allows $N_{cool}$ to be estimated. These column densities are valid for general cooling flows, which is why the analysis explains the data so well.

We interpreted subsonic cooling flows as cooling layers in a radiative shock, and supersonic cooling flows as cooling galactic winds or accretion flows, and wrote down the relationship between the shock or terminal velocity ($v_\infty$), and the temperature and cooling velocity (or line width) of the cooling flow. Thus, our model also enabled us to independently constrain the dynamics of the cooling gas by computing the shock or terminal velocities associated with such flows at the maximum likelihood temperature of the {\OVI} absorbers. We confirmed the existence of both subsonic and supersonic cooling flows in the observations. The absorption structures we observe associated with  shocks are likely two dimensional sheet like structures. The absorbers associated with supersonic cooling flows are mostly down-the-barrel observations of galactic winds and of the CGM around star-forming galaxies.

Our picture is that the CGM is a complex non-equilibrium system with a fractal like multiphase medium and with strong entropy flows due to heating, cooling and other feedback processes. Despite such complexities, we have invoked simple idealized calculations to gain useful insight into the formation and dynamics of highly ionized gas. The merit of the simple model presented here is that in general the observations of  {\OVI} and similar highly ionized absorption lines can be explained by assuming that they arise in collisionally ionized gas with flow speeds of order the speed of sound. The implications are that the data are pointing to cooling flows and shocks (see Figure~\ref{v_shock_delta_v}), including cooling feedback outflows and cooling accretion inflows and shocks where the gas is returning to the galaxy in the general CGM circulation. Our analysis also gives a simple physical picture that modelers can incorporate and that observers can use in their analysis.  The future challenge will be to undertake large surveys of higher ionization species (such as {\NeVIII} and {\MgX} )in absorption and deduce constraints on galaxy formation and CGM gas flows.  Simultaneously, the next generation of integral field spectrographs such as MUSE and KCWI, will allow imaging of the gas structures we have inferred from the absorption line data.  The next step is to incorporate these new data systematically into this type of analysis and then design better and cleaner observational diagnostics.

\section{Acknowledgment} 
Support for this work was provided by NASA through Hubble Fellowship grant \#51354 awarded by the Space Telescope Science Institute, which is operated by the Association of Universities for Research in Astronomy, Inc., for NASA, under contract NAS 5-26555. AYW has been supported in part by ERC Project No. 267117 (DARK) hosted by Universit\'e Pierre et Marie Curie (UPMC) - Paris 6, PI J.  Silk. This paper has greatly benefited from discussions with Guangtun Ben Zhu, Claudia Scarlata, Andrew Fox. We thank Chris McKee for a critical reading of the manuscript, which lead to many improvements.  We also thank David Weinberg for his comments and suggestions which helped improve the paper. C.N. particularly acknowledges very many stimulating Galaxy Journal Club Friday lunchtime discussions with expert colleagues at STScI.

\renewcommand\bibsection{}
\bibliographystyle{thesis_bibtex}
\bibliography{mybibliography}

\appendix

\setcounter{figure}{0} \renewcommand{\thefigure}{A.\arabic{figure}} 
\begin{figure}
\centering
\includegraphics[height=8cm,width=6cm]{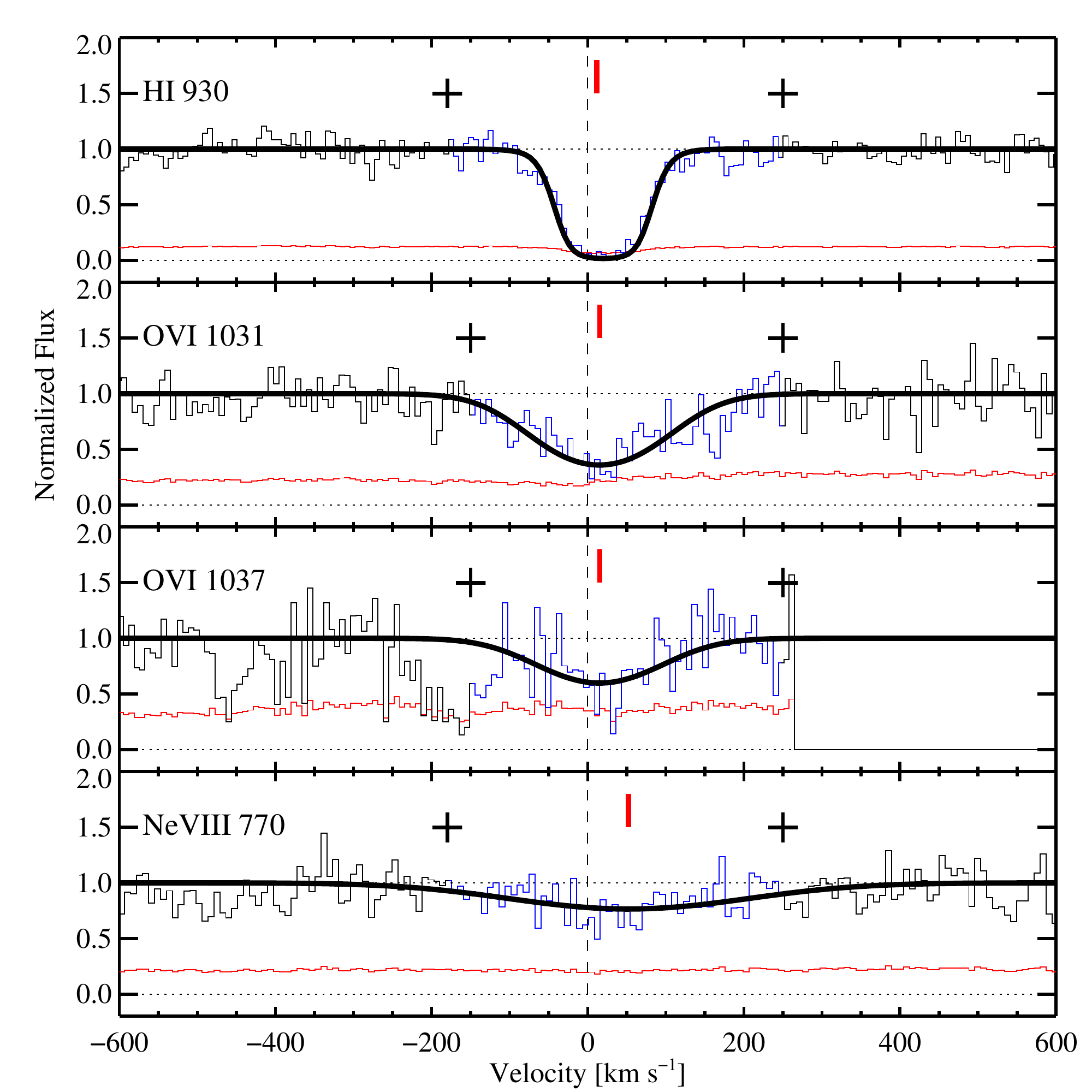}
  \caption{An additional HST-COS quasar spectra of the {\HI}, {\OVI} and {\NeVIII} absorption lines with their corresponding Voigt profile fits (solid black line) are shown, which are included in this study. The vertical red ticks indicate the centroids of individual Voigt profile components and the black crosses show the velocity range over which the profile was integrated to compute their equivalent widths. }
\label{fig:appendix}
\end{figure}

\section{COS Spectrum}
The HST-COS spectrum of a pair of new {\NeVIII} and {\OVI} absorption lines and the associated Voigt profile fits are shown in Figure \ref{fig:appendix} . This was observed along QSO J1154+4635 under HST PID 13852 (PI: Bordoloi). The Voigt profile fitted column densities and line widths for {\OVI}  are $\log\; N_{OVI}$/{\pcm} = 14.71$\pm$0.03, and $b_{OVI}$ = 99 $\pm$ 11 {\kms}; for {\NeVIII} are $\log\; N_{NeVIII}$/{\pcm} = 14.65$\pm$0.08, and $b_{NeVIII}$ = 187 $\pm$ 39 {\kms}; and for {\HI} are  $\log\; N_{HI}$/{\pcm} = 16.73$\pm$0.02, and $b_{HI}$ = 39 $\pm$ 1 {\kms}, respectively.

\begin{figure*}
\includegraphics[height=4.5cm,width=6.5cm]{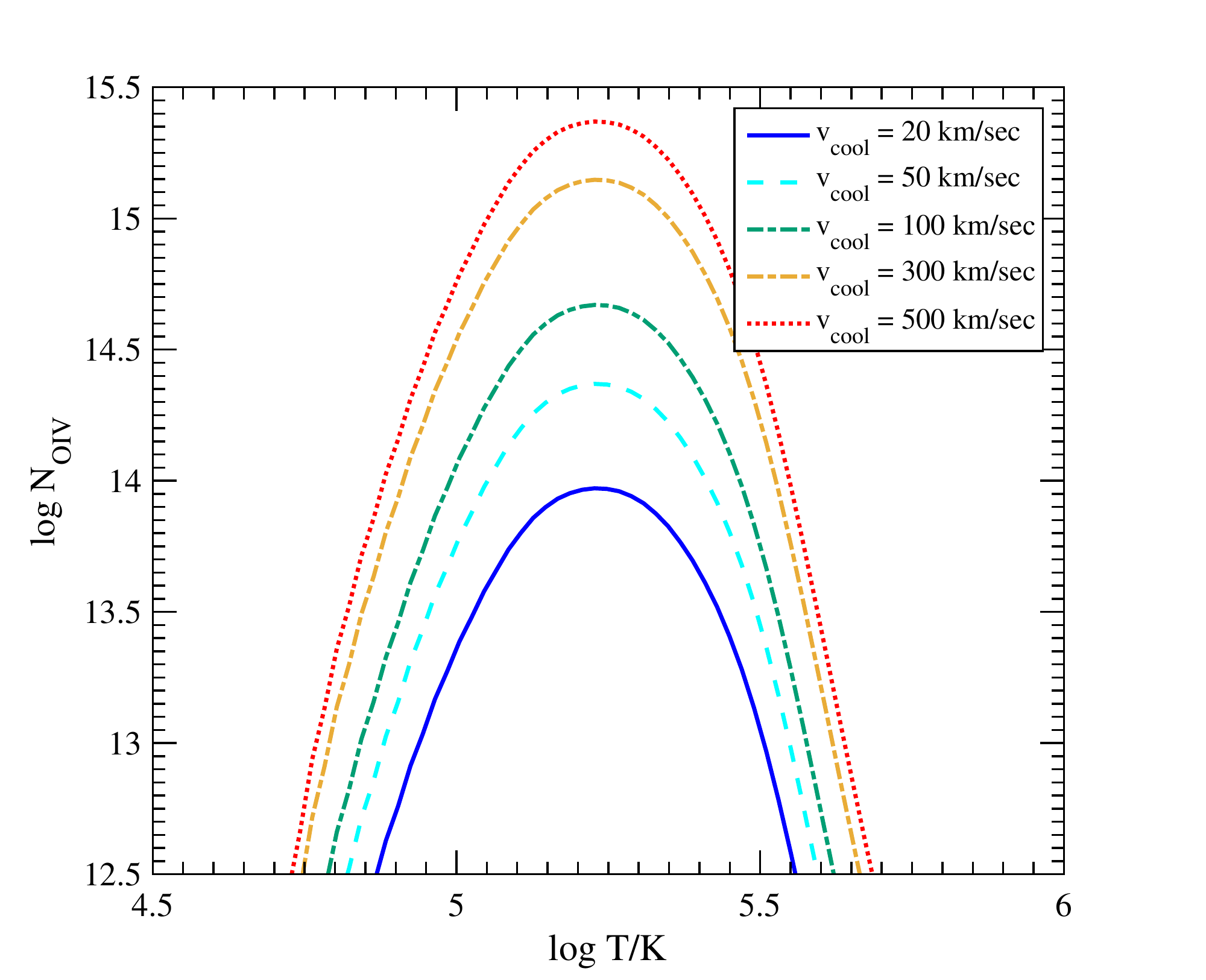}
\includegraphics[height=4.5cm,width=6.5cm]{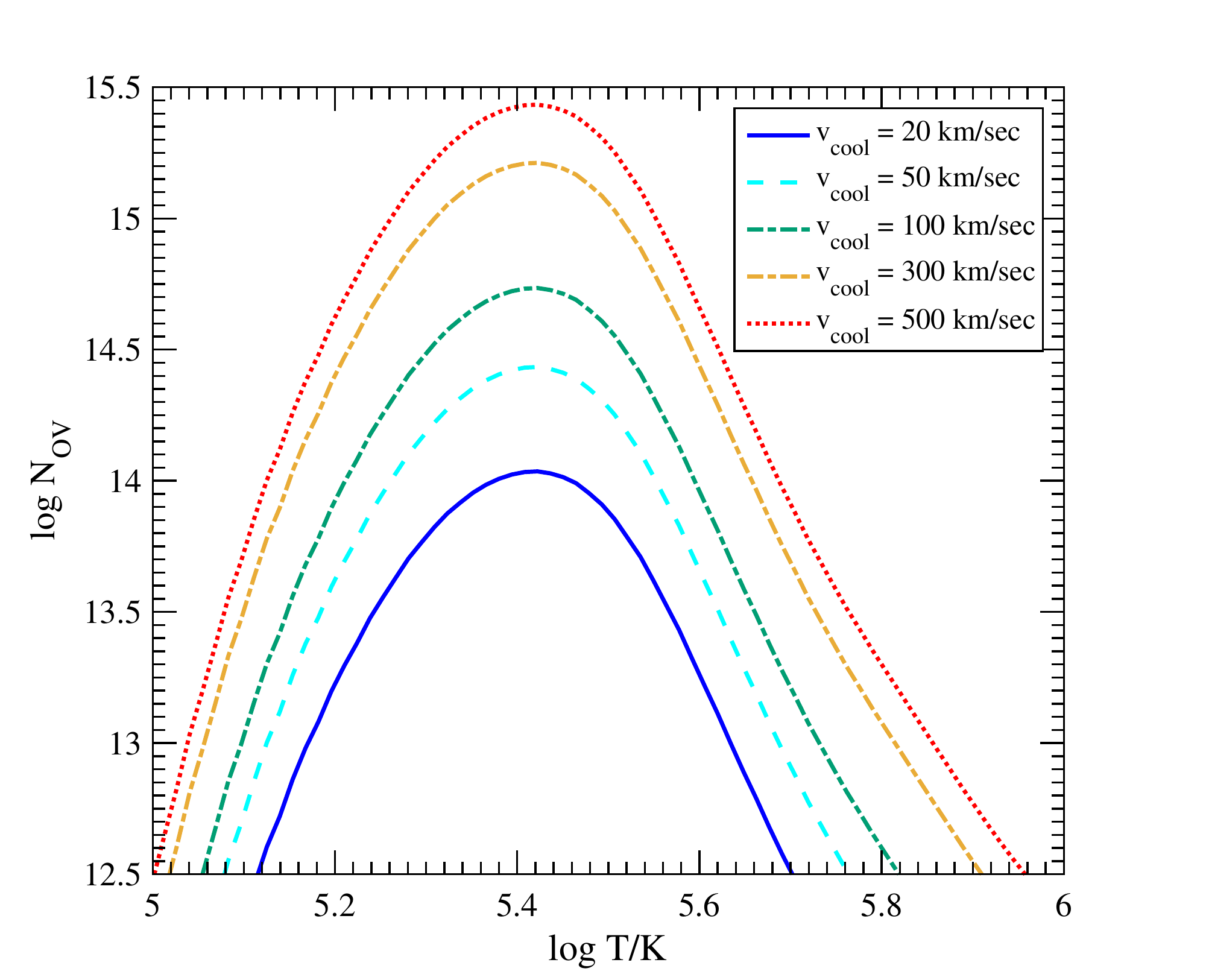}
\includegraphics[height=4.5cm,width=6.5cm]{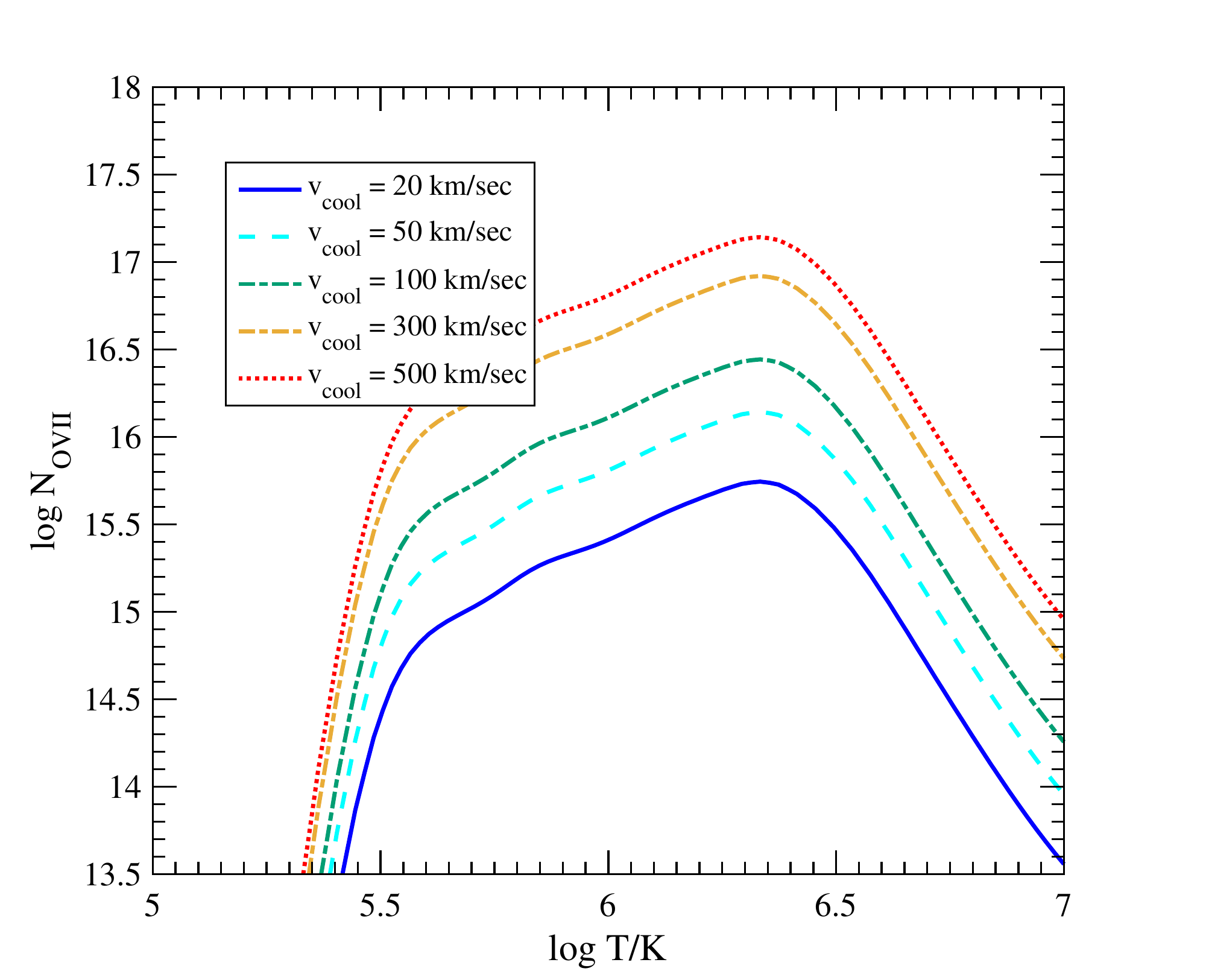}
\includegraphics[height=4.5cm,width=6.5cm]{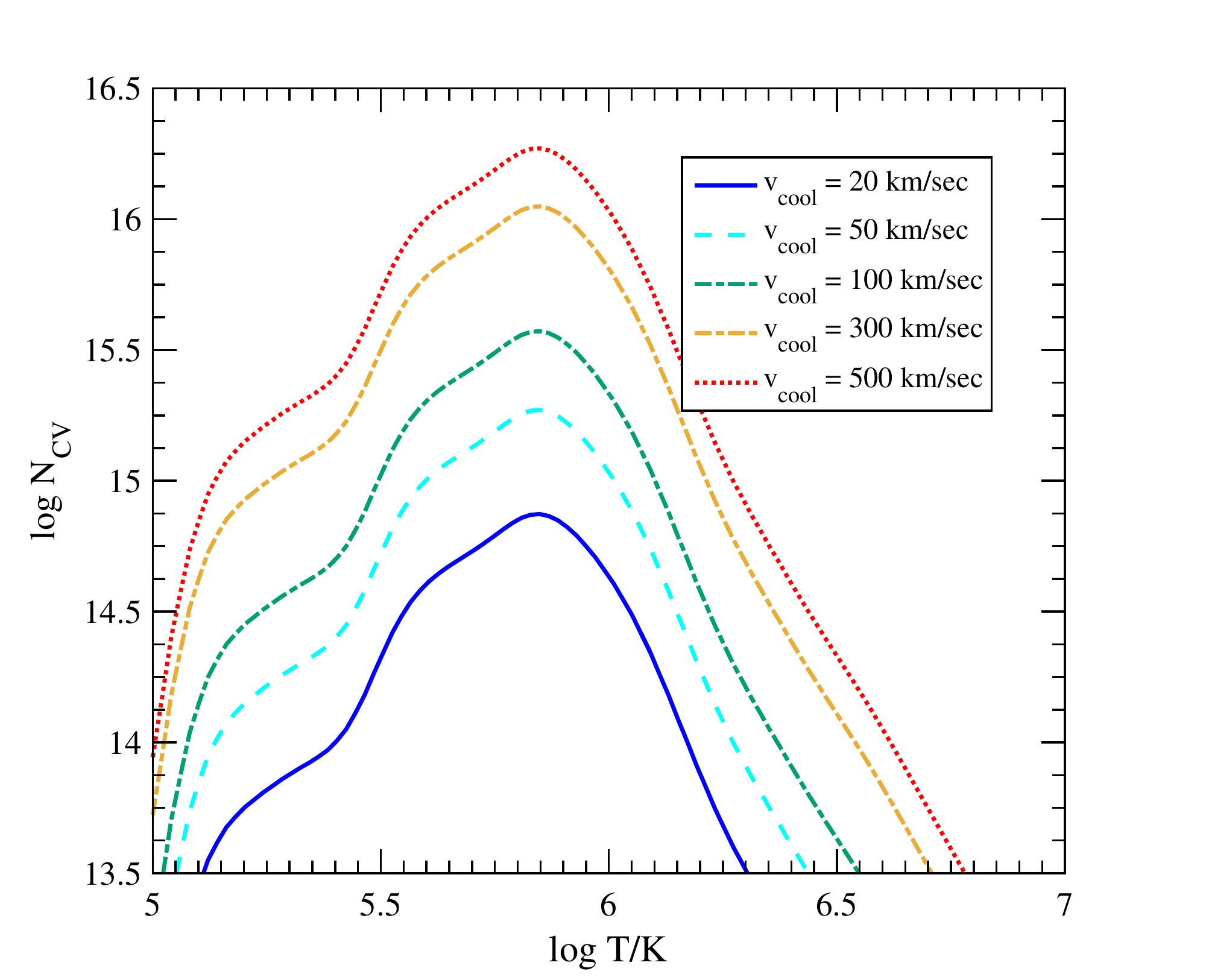}
\includegraphics[height=4.5cm,width=6.5cm]{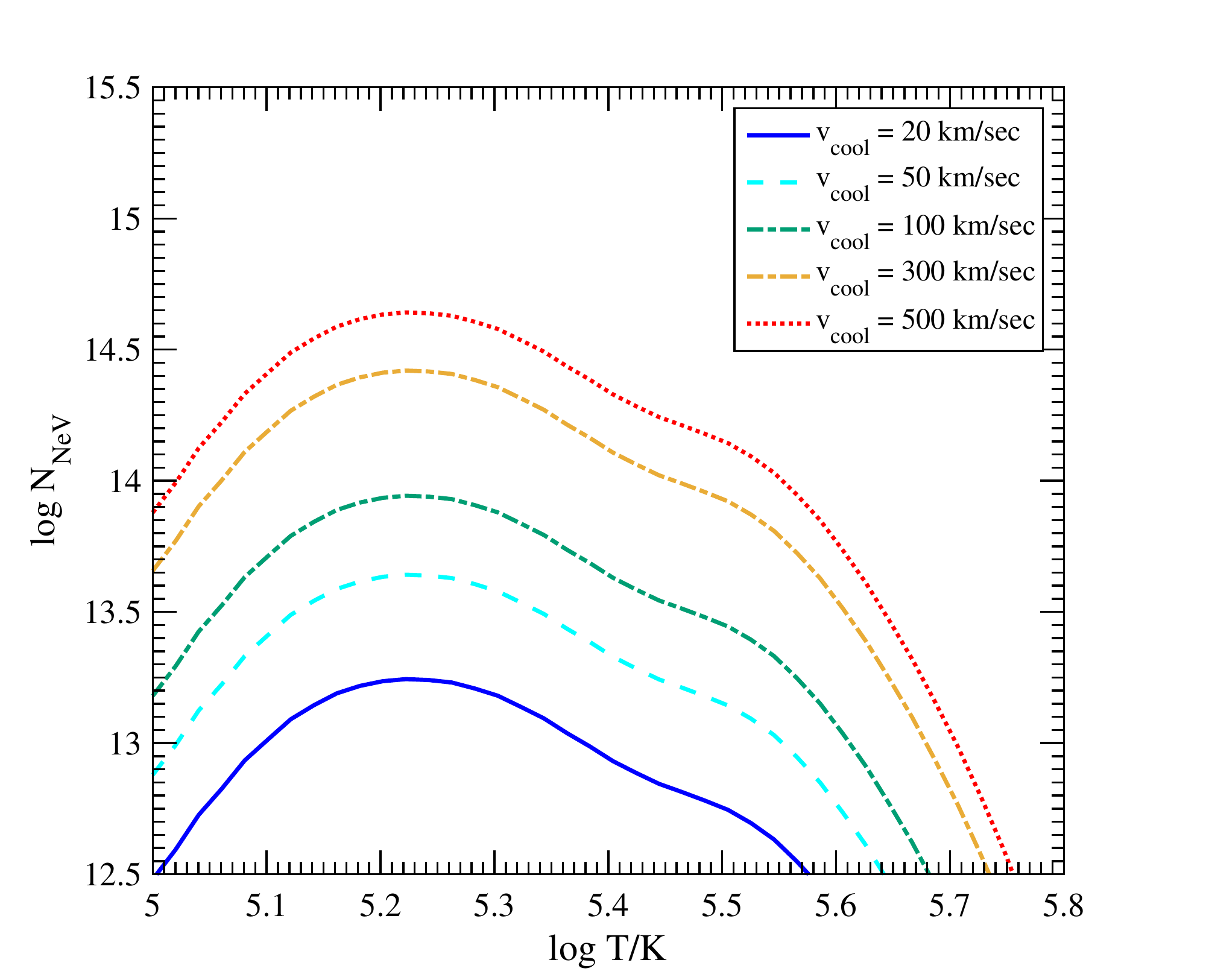}
\includegraphics[height=4.5cm,width=6.5cm]{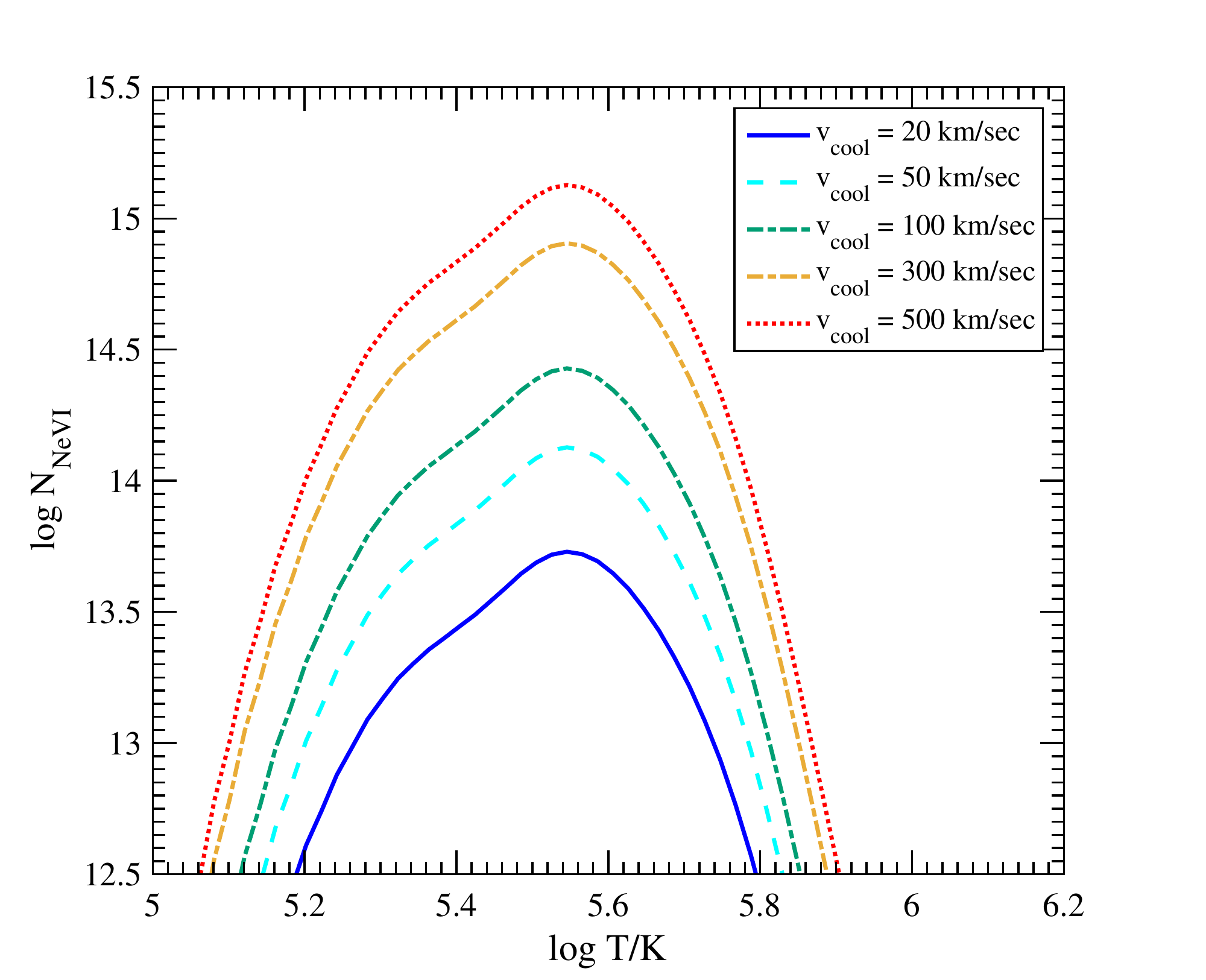}
\includegraphics[height=4.5cm,width=6.5cm]{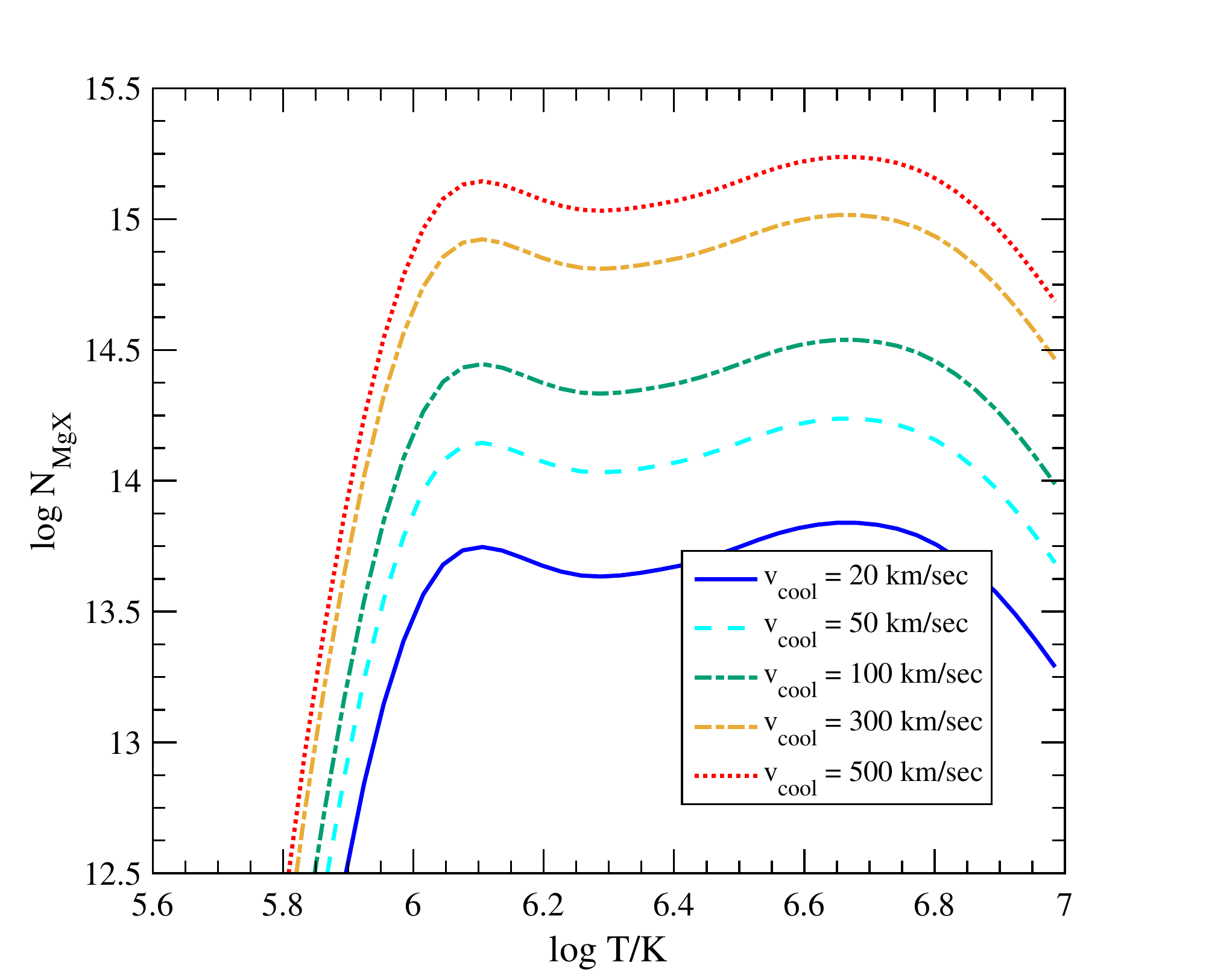}
\includegraphics[height=4.5cm,width=6.5cm]{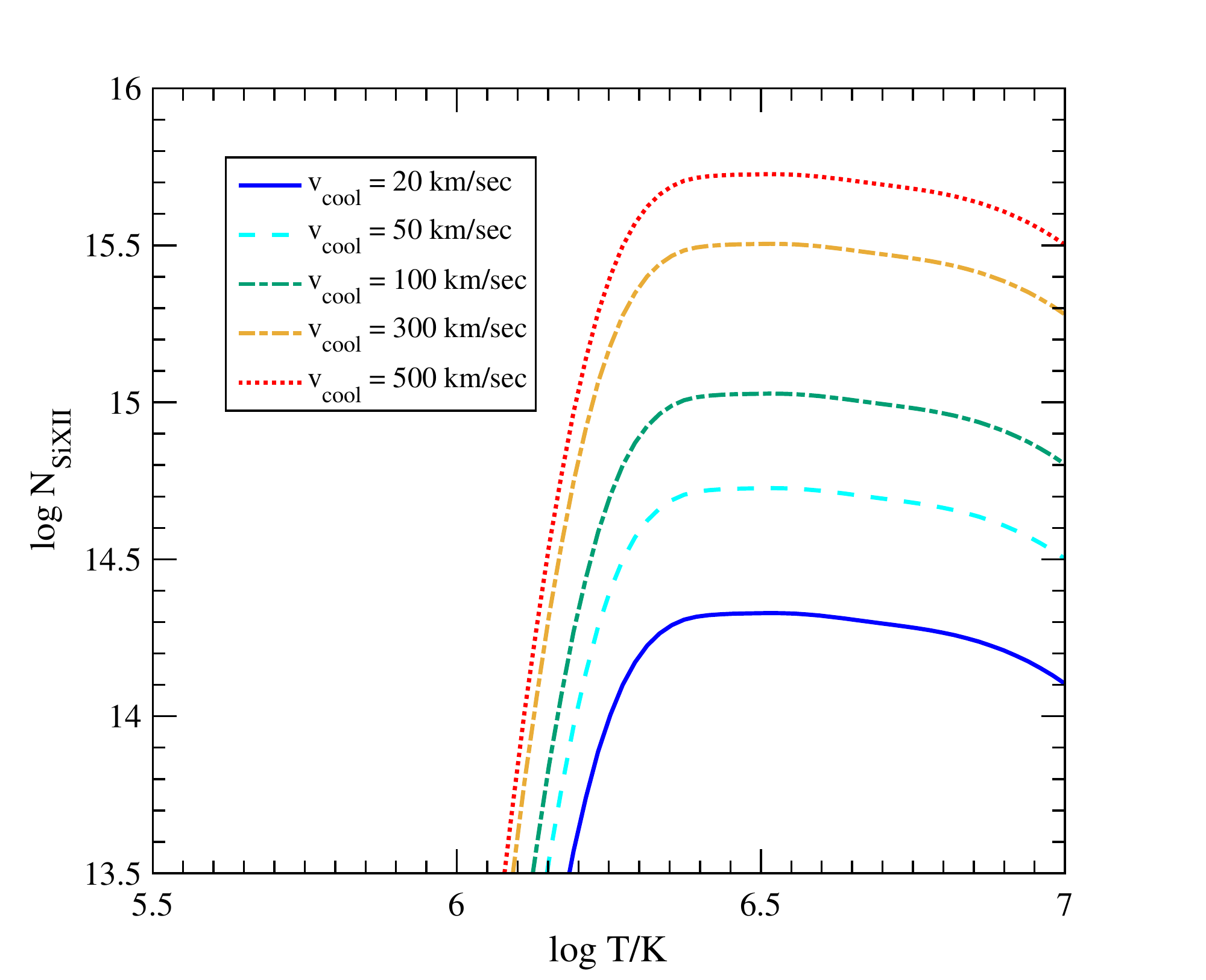}
  \caption{Cooling flow column density estimates for various ionized species as a function of their corresponding cooling temperature. Different lines represent different cooling flow velocities. }
\label{fig:predictions}
\end{figure*}

\section{Cooling flow column density predictions for several species}
In Figure \ref{fig:predictions}, we present the cooling flow model column density, temperature predictions for several species for which we do not compare the model to the data.

\end{document}